\documentclass[twoside,11pt]{article}

\usepackage[accepted]{melba}

\usepackage{amsmath,amsfonts}
\usepackage{subfiles}
\usepackage{amssymb}
\usepackage{multirow}
\usepackage{graphicx}
\usepackage[caption=false]{subfig}
\usepackage{mathrsfs}
\usepackage[dvipsnames]{xcolor}
\melbaheading{2022:001}{https://www.melba-journal.org/article/2022:001}{2022}{1-58}{09/2021}{02/2022}{Pal and Rathi}{}{}

\ShortHeadings{DL Methods for MRI Reconstruction}{Pal and Rathi}
\firstpageno{1}

\title{A review and experimental evaluation of deep learning methods for MRI reconstruction}

\author{\name Arghya Pal \email apal5@bwh.harvard.edu \\  % start right after \author{, or there will be an extra space
	\addr Department of Psychiatry and Radiology, Harvard Medical School, Boston, MA, USA
	\AND
	\name Yogesh Rathi \email yogesh@bwh.harvard.edu \\
	\addr Department of Psychiatry and Radiology, Harvard Medical School, Boston, MA, USA
}

\newcommand{\generated}{$\textbf{x}^{gen}$}

\newcommand{\loss}{\mathcal{L}}
\begin{document}
\maketitle
\vspace{0.3cm}

% abstract
\begin{abstract}%
Following the success of deep learning in a wide range of applications, neural network-based machine-learning techniques have received significant interest for accelerating magnetic resonance imaging (MRI) acquisition and reconstruction strategies. A number of ideas inspired by deep learning techniques for computer vision and image processing have been successfully applied to nonlinear image reconstruction in the spirit of compressed sensing for accelerated MRI. Given the rapidly growing nature of the field, it is imperative to consolidate and summarize the large number of deep learning methods that have been reported in the literature, to obtain a better understanding of the field in general. This article provides an overview of the recent developments in neural-network based approaches that have been proposed specifically for improving parallel imaging. A general background and introduction to parallel MRI is also given from a classical view of k-space based reconstruction methods. Image domain based techniques that introduce improved regularizers are covered along with k-space based methods which focus on better interpolation strategies using neural networks. While the field is rapidly evolving with plenty of papers published each year, in this review, we attempt to cover broad categories of methods that have shown good performance on publicly available data sets. Limitations and open problems are also discussed and recent efforts for producing open data sets and benchmarks for the community are examined.
\end{abstract}

% keywords

\section{Introduction}
% Paragraph 1:
    % Introduction to the MRI
    % The problems associated with MRI
    % Classical methods
Magnetic Resonance Imaging (MRI) is an indispensable clinical and research tool used to diagnose and study several diseases of the human body. It has become a \emph{sine qua non} in various fields of radiology, medicine, and psychiatry. 
Unlike computed tomography (CT), it can provide detailed images of the soft tissue and does not require any radiation, thus making it less risky to the subjects. 
MRI scanners sample a patient's anatomy in the frequency domain that we will call ``k-space". 
The number of rows/columns that are acquired in k-space is proportional to the quality (and spatial resolution) of the reconstructed MR image. 
To get higher spatial resolution, longer scan time is required due to the increased number of k-space points that need to be sampled \citep{fessler2010model}. 
Hence, the subject has to stay still in the MRI scanner for the duration of the scan to avoid signal drops and motion artifacts. 
Many researchers have been trying to reduce the number of k-space lines to save scanning time, which leads to the well-known problem of ``aliasing'' as a result of the violation of the Nyquist sampling criteria \citep{nyquist1928certain}. Reconstructing high-resolution MR images from the undersampled or corrupted measurements was a primary focus of various sparsity promoting methods, wavelet-based methods, edge-preserving methods, and low-rank based methods. 
This paper reviews the literature on solving the inverse problem of MR image reconstruction from noisy measurements using Deep Learning (DL) methods, while providing a brief introduction to classical optimization based methods. We shall discuss more about this in Sec. \ref{sec_intro_mathematical_formulaion}. 

% Paragraph 2
    % We will be talking about DL
    % The successes of DL in other domains
A DL method learns a non-linear function $f:\mathcal{Y}\rightarrow\mathcal{X}$ from a set of all possible mapping functions $\mathcal{F}$. 
The accuracy of the mapping function can be measured using some notion of a loss function $l: \mathcal{Y} \times \mathcal{X}\rightarrow[0, \infty)$. 
The empirical risk \citep{vapnik1991principles}, $\hat{L}(f)$, can be estimated as $\hat{L}(f)=\frac{1}{2}\sum_{i=1}^{m}l(f(\textbf{y}_i), \textbf{x}_i)$ and the generalization error of a mapping function $f(\cdot)$ can be measured using some notion of accuracy measurement. 
MR image reconstruction using deep learning, in its simplest form, amounts to learning a map $f$ from the undersampled k-space measurement $\mathcal{Y}\in \mathbb{C}^{N_1\times N_2}$, or $\mathcal{Y}\in \mathbb{R}^{N_1\times N_2\times 2}$  to an unaliased MR image $\mathcal{X}\in \mathbb{C}^{N_1\times N_2}$,  or $\mathcal{Y}\in \mathbb{R}^{N_1\times N_2\times 2}$, where $N_{1}$, $N_{2}$ are the height and width of the complex valued image. 
In several real-world cases, higher dimensions such as time, volume, etc., are obtained and accordingly the superscripts of $\mathcal{Y}$ and $\mathcal{X}$ change to $\mathbb{C}^{N_1\times N_2 \times N_3\times N_4 \times \cdots}$. 
For the sake of simplicity, we will use assume $\mathcal{Y}\in \mathbb{C}^{N_1\times N_2}$ and $\mathcal{X}\in \mathbb{C}^{N_1\times N_2}$.

%The DL-based methods have shown thriving successes in different facets of human lives - from the morning news feed, the shortest path to our maps, apparel selection, and foreign language understanding. 
%The omnipresence of DL methods are becoming more visible due to the success of AlphaGo \citep{alphago}, Generative Pre-trained Transformers (GPTs) \citep{gpt}, BigGAN \citep{biggan} and other methods those collectively excelling in different dimensions, such as recognition, generation, explanation, and imagination. 
In this survey, we focus on two broad aspects of DL methods, i.e. (i) generative models, which are data generation processes capturing the underlying density of data distribution; 
and (ii) non-generative models, that learn complex feature representations of images intending to learn the inverse mapping from k-space measurements to MR images. 
%In each passing day, DL methods in Computer Vision and MR reconstruction domain are moving one more step towards the goal of mitigating the gap between artificial cognition and human-like cognition using more advanced algorithms. 
Given the availability and relatively broad access to open-source platforms like Github, PyTorch \citep{Pytorch}, and TensorFlow \citep{tensorflow}, as well as large curated datasets and high-performance GPUs, deep learning methods are actively being pursued for solving the MR image reconstruction problem with reduced number of samples while avoiding artifacts and boosting the  signal-to-noise ratio (SNR).
%%%%%%%%%%%%%%%%%%%%%
%%%%%%%%%%%%%%%%
%%%%%%%%%%
%%%%
%
\begin{figure*}[!t]
    \centering
    \includegraphics[width=\textwidth]{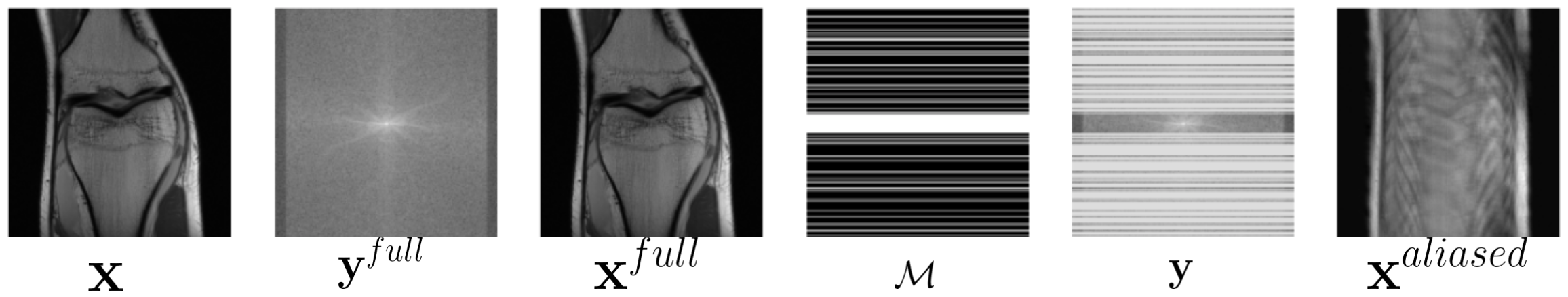}
    \caption{(\textbf{Left to Right}) An example of a fastMRI \citep{zbontar2018fastmri} knee image $\textbf{x}$, fully sampled k-space $\textbf{y}^{full}$, corresponding reconstructed image $\textbf{x}^{full}$ from $\textbf{y}^{full}$, the sampling mask $\mathcal{M}$ of fastMRI that we apply on fully sampled k-space $\textbf{y}^{full}$, the sampled k-space $\textbf{y}$, and the corresponding aliased reconstructed image $\textbf{x}^{aliased}$.}
    \label{fig_intro_1}
\end{figure*}
%%%%%%%%%%%%%%%%%%%%%
%%%%%%%%%%%%%%%
%%%%%%%
%%
%

In Sec. \ref{sec_intro_mathematical_formulaion}, we briefly discuss the mathematical formulation that utilizes k-space measurements from multiple receiver coils to reconstruct an MR image. 
Furthermore, we discuss some challenges of the current reconstruction pipeline and discuss the DL methods (in Sec. \ref{sec_intro_deep_learning}) that have been introduced to address these limitations. We finally discuss the open questions and challenges to deep learning methods for MR reconstruction in sections \ref{sec_classical_k_space_methods}, \ref{sec_classic_image}, and \ref{sec_DLBB}. 
%%%%%%%%%%%%%%%%%%%
%%%%%%%%%%%%%%
%%%
%
\subsection{Mathematical Formulation for Image Reconstruction in Multi-coil MRI}
\label{sec_intro_mathematical_formulaion}
%%%%%%%%%%%%%%%
Before discussing undersampling and the associated aliasing problem, let us first discuss the simple case of reconstructing an MR image, $\textbf{x}\in \mathbb{C}^{N_{1}\times N_{2}}$, from a fully sampled k-space measurement, $\textbf{y}^{full}\in\mathbb{C}^{N_{1}\times N_{2}}$, using the Fourier transform $\mathscr{F}(\cdot)$:
\begin{equation}
    \textbf{y}^{full} = \mathscr{F}\textbf{x} + \eta,
    \label{eq_basic}
\end{equation}
where $\eta \sim \mathcal{N}(0, \Sigma)$ is the associated measurement noise typically assumed to have a Gaussian distribution \citep{virtue2017empirical} when the k-space measurement is obtained from a single receiver coil. 

Modern MR scanners support parallel acquisition using an array of overlapping receiver coils $n$ modulated by their sensitivities $S_i$. So Eqn. \ref{eq_basic} changes to: $\textbf{y}_{i}^{full} = \mathscr{F}S_{i}\textbf{x} + \eta$, where $i=\{1, 2, \cdots, n\}$ is the number of receiver coils. 
We use $\textbf{y}_{i}$ for the undersampled k-space measurement coming from the $i^{th}$ receiver coil. 
To speed up the data acquisition process, multiple lines of k-space data (for cartesian sampling) are skipped using a binary sampling mask $\mathcal{M}\in\mathbb{C}^{N_{1}\times N_{2}}$ that selects a subset of k-space lines from $\textbf{y}^{full}$ in the phase encoding direction:
%%%%%%%%%%%%%%%%%%%%%
\begin{equation}
    \textbf{y}_{i} = \mathcal{M}\odot\mathscr{F}S_{i}\textbf{x} + \eta.
\label{eq_1_cs_mri}
\end{equation}
%%%%%%%%%%%%%%%%%%%%%%
An example of $\textbf{y}^{full}$, $\textbf{y}$, $\mathcal{M}$ is shown in Fig \ref{fig_intro_1}. 

To estimate the MR image $\textbf{x}$ from the measurement, a data fidelity loss function is typically used to ensure that the estimated data is as close to the measurement as possible. 
A typical loss function is the squared loss, which is minimized to estimate $\textbf{x}$:
\begin{equation}
\begin{split}
    \hat{\textbf{x}} = \arg\min_{\textbf{x}} \frac{1}{2}\sum_{i}||\textbf{y}_{i}-\mathcal{M}\odot\mathscr{F}S_{i}\textbf{x}||^{2}_{2} = \arg\min_{\textbf{x}} ||\textbf{y}-A\textbf{x}||^{2}_{2}.
\end{split}
\label{eq_2_least_square}
\end{equation}
We borrow this particular formulation from  \citep{sriram2020end, zheng2019cascaded}. 
This squared loss function is quite convenient if we wish to compute the error gradient during optimization. 

However, the formulation in Eqn. \ref{eq_2_least_square} is under-determined if data is undersampled and does not have a unique solution. Consequently, a regularizer $\mathcal{R}(\textbf{x})$ is typically added to solve such an ill-conditioned cost function:
\footnote{The regularization term, $\mathcal{R}(\textbf{x})$ is related to the prior, $p(\textbf{x})$, of a maximum \textit{a priori} (MAP) extimation of $\textbf{x}$, i.e. $\hat{\textbf{x}} = \arg \min_{\textbf{x}}(-\log p(\textbf{y}|\textbf{x})-\log p(\textbf{x}))$. 
In fact, in \cite{ravishankar2019image} the authors loosely define $\beta \mathcal{R}(\textbf{x}) = -\log p(\textbf{x})$, which promotes some desirable image properties such as spatial smoothness, sparsity in image space, edge preservation, etc. with a view to get a unique solution.
}
\begin{equation}
    \hat{\textbf{x}} = \arg\min_{\textbf{x}}\frac{1}{2}||\textbf{y} - A\textbf{x}||^{2}_{2} + \sum_{i} \lambda_{i}\mathcal{R}_{i}(\textbf{x}).
\label{eq_3_tikonov}
\end{equation}
Please note that each $\mathcal{R}_{i}(\textbf{x})$ is a separate regularizer, while the $\lambda_{i}$s are hyperparameters that control the properties of the reconstructed image $\hat{\textbf{x}}$ while avoiding over-fitting. 
Eqn. \ref{eq_2_least_square} along with the regularization term can be optimized using various methods, such as 
(i) the Morozov formulation, $\hat{\textbf{x}} = \min\{\mathcal{R}(\textbf{x}); \text{such that}, ||A\textbf{x} -\textbf{y}||\leq \delta\}$; 
(ii) the Ivanov formulation, i.e. $\hat{\textbf{x}} = \min\{||A\textbf{x} -\textbf{y}||; \text{such that}, \mathcal{R}(\textbf{x})\leq \epsilon\}$; or 
(iii) the Tikonov formulation, $\hat{\textbf{x}} = \min\{||A\textbf{x} -\textbf{y}|| + \lambda\mathcal{R}(\textbf{x})\}$, discussed in   \citep{oneto2016tikhonov}. 
%In this study, we consider the Tikonov formulation that is quite famous for its closed form solution. 

In general, the Tikonov formulation can be designed using a physics based, sparsity promoting, dictionary learning, or a deep learning based model. But there are several factors that can cause loss in data quality (especially small anatomical details) such as inaccurate modeling of the system noise, complexity, generalizability etc. To overcome these limitations, it is essential to develop inverse mapping methods that not only provide good data fidelity but also generalize well to unseen and unexpected data. In the next section, we shall describe how DL methods can be used as priors or regularizers for MR reconstruction.
%%%%%%%%%%%%%%%%%%%%
%%%%%%%%%%%%%%%
%%%
%
\begin{figure*}[!t]
    \centering
    \includegraphics[width=\textwidth]{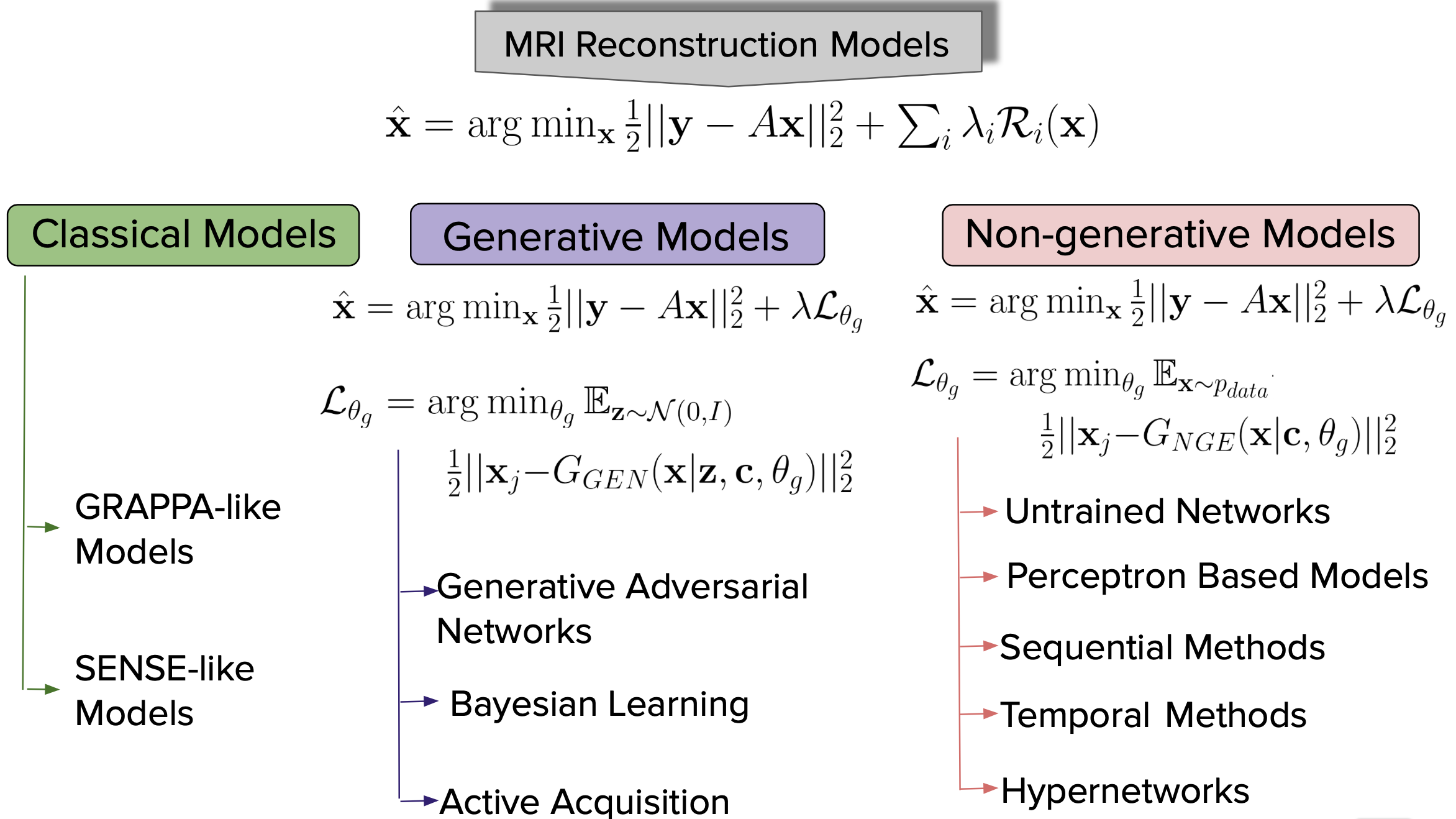}
    \caption{\textbf{MRI Reconstruction Methods}: We shall discuss various MR reconstruction methods with a main focus on Deep Learning (DL) based methods. Depending on the optimization function, a DL method can be classified into a Generative (discussed in Sec. \ref{sec_generative_model}) or a Non-Generative model (discussed in Sec. \ref{sec_nongen}). However, for the sake of completeness, we shall also discuss classical k-space (GRAPPA-like) and image space based (SENSE-like) methods.}.
    \label{fig_intro_2}
\end{figure*}
%%%%%%%%%%%%%%%%%%%%%
%%%%%%%%%%%%%%%
%%%%%%%
%%
%
\subsection{Deep Learning Priors for MR Reconstruction}
\label{sec_intro_deep_learning}
We begin our discussion by considering DL methods with learn-able parameters $\theta$. 
The learn-able parameters $\theta$ can be trained using some notion of a learning rule that we shall discuss in Sec. \ref{sec_DLBB}. 
A DL training process helps us to find a function $G_{DL}(\cdot)$ that acts as a regularizer to Eqn. \ref{eq_3_tikonov} with an overarching concept of an inverse mapping, i.e; \big(please note that, we shall follow \citep{zheng2019cascaded} to develop the optimization formulation\big)
\begin{equation}
\begin{split}
    \hat{\textbf{x}} = \arg\min_{\textbf{x}}\frac{1}{2}||\textbf{y} - A\textbf{x}||^{2}_{2} + \lambda\mathcal{L}_{\theta}; \text{where}, \mathcal{L}_{\theta} = \arg\min_{\theta}\sum_{j}||\textbf{x}_{j} - G_{DL}(\textbf{x}|\textbf{z},\textbf{c},\theta)||^{2}_{2}
\end{split}
\label{eq_3_deep_learning}
\end{equation}
and $\textbf{z}$ is a latent variable capturing the statistical regularity of the data samples, while $\textbf{c}$ is a conditional random variable that depends on a number of factors such as: undersampling of the k-space  \citep{SUBGAN, AutomapGAN, AMRI}, the resolution of the image \citep{DAGAN, SARAGAN}, or the type of DL network used \citep{COLLAGAN}. %which determine the type of probabilistic mapping function we wish to learn. 
Based on the nature of the learning, there are two types of DL methods known as  \textit{generative models}, and \textit{non-generative models}. 
We shall start with a basic understanding of DL methods to a more in-depth study of different architectures in Secs. \ref{sec_generative_model} and \ref{sec_nongen}. 

In generative modeling, the random variable $\textbf{z}$ is typically sampled from a noisy Gaussian distribution, $\textbf{z}\sim\mathcal{N}(0, I)$ with or without the presence of the conditional random variable, i.e.;
\begin{equation}
\begin{split}
    \hat{\textbf{x}} = \arg\min_{\textbf{x}}\frac{1}{2}||\textbf{y} - A\textbf{x}||^{2}_{2} + \lambda\mathcal{L}_{\theta_g}; \mathcal{L}_{\theta_g} = \arg\min_{\theta_{g}}\mathbb{E}_{\textbf{z}\sim\mathcal{N}(0,I)}\frac{1}{2}||\textbf{x}_{j} - G_{GEN}(\textbf{x}|\textbf{z},\textbf{c},\theta_{g})||^{2}_{2}
\end{split}
\label{eq_4_generative_models}
\end{equation}
There are various ways to learn the parameters of Eqn. \ref{eq_4_generative_models}. 
For instance, the Generative Adversarial Network (GAN) \citep{GAN} allows us to learn the generator function $G_{GEN}(\cdot)$ using an interplay between two modules, while the Variational Autoencoders (VAEs) \citep{vae} learns $G_{GEN}(\cdot)$ by optimizing the evidence lower bound (ELBO), or by incorporating a prior in a Bayesian Learning setup as described in Section \ref{sec_vaes}. 
It is shown in the literature that a generative model can efficiently de-alias an MR image that had undergone a $4\times$ or $8\times$ undersampling in k-space \citep{zbontar2018fastmri}.

The non-generative models on the other hand do not learn the underlying latent representation, but instead learn a mapping from the measurement space to the image space. 
Hence, the random variable $\textbf{z}$ is not required. 
The cost function for a non-generative model is given by:
%{\textbf{REMOVE z from the expression}}
\begin{equation}
\begin{split}
    \hat{\textbf{x}} = \arg\min_{\textbf{x}}\frac{1}{2}||\textbf{y} - A\textbf{x}||^{2}_{2} + \lambda\mathcal{L}_{\theta_g}; \mathcal{L}_{\theta_g} = \arg\min_{\theta_{g}}\mathbb{E}_{\textbf{x}\sim p_{data}}\frac{1}{2}||\textbf{x}_{j} - G_{NGE}(\textbf{x}|\textbf{c},\theta_{g})||^{2}_{2}.
\end{split}
\label{eq_5_non_generative_models}
\end{equation}
The function $G_{NGE}(\cdot)$ is a non-generative mapping function that could be a Convolutional Neural Network (CNN) \citep{zheng2019cascaded, RAKI, sriram2020grappanet}, a Long Short Term Memory (LSTM) \citep{hochreiter1997long}, or any other similar deep learning model. 
The non-generative models show a significant improvement in image reconstruction quality compared to classical methods. We shall describe the generative and the non-generative modeling based approaches in detail in Secs. \ref{sec_generative_model} and \ref{sec_nongen} respectively. Below, we give a brief overview of the classical or the non-DL based approaches for MR image reconstruction.
%%%%%%%%%%%%%%%%%%%%%
%%%%%%%%%%%%%%%
%%%%%%%
%%
%
\subsection{Classical Methods for MR Reconstruction}
\label{sec_intro_non_deep_learning}
In the literature, several approaches can be found that perform an inverse mapping to reconstruct the MR image from k-space data. 
Starting from analytic methods \citep{fessler2003nonuniform,laurette1996cone}, there are several works that provide ways to do MR reconstruction, such as the physics based image reconstruction methods \citep{roeloffs2016model, tran2016model, maier2019rapid, tran2013model, hilbert2018accelerated, sumpf2011model, ben2016accelerated, zimmermann2017accelerated, schneider2020free}, the sparsity promoting compressed sensing methods \citep{feng1996spectrum, bresler1996spectrum, candes2006robust}, and low-rank based approaches \citep{haldar2013low}.  
All these methods can be roughly categorized into two categories, i.e. 
(i) GRAPPA-like methods: where prior assumptions are imposed on the k-space; and 
(ii) SENSE-like methods: where an image is reconstructed from the k-space while jointly unaliasing (or dealiasing) the image using sparsity promoting and/or edge preserving image regularization terms. %We shall consider the division of k-space based methods and image space based method to discuss classical methods. 

A few k-space methods estimate the missing measurement lines by learning kernels from an already measured set of k-space lines from the center of the k-space (i.e., the auto-calibration or ACS lines). 
%Such kernels are linear regressors that act as priors and provide plausible solutions to the under-determined problem of MR reconstruction. 
These k-space based methods include methods such as SMASH \citep{SMASH}, VDAUTOSMASH \citep{VDAUTOSMASH}, GRAPPA and its variations \citep{bouman1993generalized, park2005artifact, seiberlich2008self}. 
The k-t GRAPPA \citep{huang2005k} takes advantage of the correlations in the k-t space and interpolates the missing data. 
On the other hand, sparsity promoting low rank based methods are based on the assumption that, when the image reconstruction follows a set of constraints (such as sparsity, smoothness, parallel imaging, etc.), the resultant k-space should follow a structure that has low rank. 
The low rank assumption has been shown to be quite successful in dynamic MRI \citep{liang2007spatiotemporal}, functional MRI \citep{singh2015under}, and diffusion MRI \citep{hu2019motion}. In this paper we give an overview of the low-rank matrix approaches \citep{haldar2013low, jin2016general, lee2016acceleration, ongie2016off, haldar2016p, haldar2017computational} in Sec. \ref{sec_classical_k_space_methods}. 
While in k-t SLR \citep{lingala2011accelerated}, a spatio-temporal total variation norm is used to recover the dynamic signal matrix.

The image space based reconstruction methods, such as, the model based image reconstruction algorithms, incorporate the underlying physics of the imaging system and leverage image priors such as neighborhood information (e.g. total-variation based sparsity, or edge preserving assumptions) during image reconstruction. 
%The MBIR algorithms are such initial methods those clearly have a data fidelity term to capture noise variance and imaging model and a regularization term through which the prior information about the image is incorporated. 
%The main advantage of the iterative method is that they allow the pipeline to model statistical invariances rather than simple geometric modeling as in classical methods. 
%We note that, the statistical models are not only provide plausible image reconstruction but provides reliability on the underlying mapping, and as a consequence, such methods got the approval from the U.S Food and Drug administration in recent past. 
Another class of works investigated the use of compressed sensing (CS) in MR reconstruction after its huge success in signal processing \citep{feng1996spectrum, bresler1996spectrum, candes2006robust}. 
Compressed sensing requires incoherent sampling and sparsity in the transform domain (Fourier, Wavelet, Ridgelet or any other basis) for nonlinear image reconstruction. 
%There are also unrolled optimization methods that decouples a standard learning procedure into two (or, more) sub-learning procedures. 
We also describe dictionary learning based approaches that are a spacial case of compressed sensing using an overcomplete dictionary. 
The methods described in \citep{gleichman2011blind, ravishankar2016data, lingala2013blind, rathi2011sparse, michailovich2011spatially} show various ways to estimate the image and the dictionary from limited measurements.
%%%%%%%%%%%%%
%%%%%%%%%%
%%%%
%
\subsection{Main Highlights of This Literature Survey}
\label{sec_intro_main_aim}
The main contributions of this paper are:
\begin{itemize}
    \item We give a holistic overview of MR reconstruction methods, that includes a family of classical k-space based image reconstruction methods as well as the latest developments using deep learning methods. 
    \item We provide a discussion of the basic DL tools such as activation functions, loss functions, and network architecture, and provide a systematic insight into generative modeling and non-generative modeling based MR image reconstruction methods and discuss the advantages and limitations of each method.
    \item We compare eleven methods that includes classical, non-DL and DL methods on fastMRI dataset and provide qualitative and quantitative results in Sec. \ref{sec_discussion}.
    \item We conclude the paper with a discussion on the open issues for the adoption of deep learning methods for MR reconstruction and the potential directions for improving the current state-of-the-art methods. 
\end{itemize}

\section{Classical Methods for Parallel Imaging}
\label{sec_review_of_non_dl_methods}
This section reviews some of the classical k-space based MR image reconstruction methods and the classical image space based MR image reconstruction methods. 
%%%%%%%%%%%%%%%%%%%%%%
%%%%%%%%%%%%%%
%%%%%%%%%%%%
%%%%%%
%%
%
\subsection{Inverse Mapping using k-space Interpolation}
\label{sec_classical_k_space_methods}
Classical k-space based reconstruction methods are largely based on the premise that the missing k-space lines can be interpolated (or extrapolated) based on a weighted combination of all acquired k-space measurement lines. 
For example, in the SMASH \citep{SMASH} method, the missing k-space lines are estimated using the spatial harmonics of order $m$. The k-space signal can then be written as:
\begin{equation}
\begin{split}
\textbf{y}(k_1,k_2) = \sum_{i=1}^{n}w_{i}^{m}\textbf{y}_{i}(k_1,k_2)= \int\int dn_1dn_2\sum_{i=1}^{n}w_{i}^{m}S_{i}\textbf{x}e^{-jk_1n_1 - j(k_2 + m\Delta k_2)n_2}    
\end{split}
\label{eq_9_SMASH_equation}
\end{equation}
%%%%%%%%%%%%%%%%
%%%%%%%%%%
%%
%
where, $w_i^{m}$ are the spatial harmonics of order $m$, $\Delta k_2 = \frac{2\pi}{FOV}$ is the minimum k-space interval ($FOV$ stands for field-of-view), and $\textbf{y}(k_1, k_2)$ is the k-space measurement of an image $\textbf{x}$, $(k_1, k_2)$ are the co-ordinates in k-space along the phase encoding (PE) and frequency encoding (FE) directions, and $j$ represents the imaginary number. 
From Eqn. \ref{eq_9_SMASH_equation}, one can note that the $m^{th}$-line of k-space can be generated using $m$-number of spatial harmonics and hence we can estimate convolution kernels to approximate the missing k-space lines from the acquired k-space lines. 
So, in SMASH a k-space measurement $\textbf{x}$, also known as a composite signal in common parlance, is basically a linear combination of $n$ number of component signals (k-space measurements) coming from $n$ receiver coils modulated by their sensitivities $S_{i}$, i.e. 
%$\textbf{y} = \sum_{i=1}^{n}\mathcal{M}\odot\mathscr{F}S_{i}\textbf{x} + \eta$. 
%It is more meaningful to represent a composite signal as a weighted combination of component signals:
\begin{equation}
    \textbf{y} = \sum_{i=1}^{n}w_{i}\mathcal{M}\odot\mathscr{F}\mathcal{S}_{i}\textbf{x} + \eta.
    \label{eq_8_weighted_composite_signal}
\end{equation}
We borrow the mathematical notation from \citep{SMASH} and represent the composite signal in Eqn. \ref{eq_8_weighted_composite_signal} as follows:
\begin{equation}
\begin{split}
    \textbf{y}(k_1,k_2) = \sum_{i=1}^{n}w_{i}\textbf{y}_{i}(k_1,k_2)= \int\int dn_1dn_2\sum_{i=1}^{n}w_{i}\mathcal{M}\mathcal{S}_{i}\textbf{x}e^{-jk_1n_1-jk_2n_2}.
\end{split}
\label{eq_9}
\end{equation}
However, SMASH requires the exact estimation of the sensitivity of the receiver coils to accurately solve the reconstruction problem.

To address this limitation, AUTO-SMASH \citep{AUTOSMASH} assumed the existence of a fully sampled block of k-space lines called \textit{autocalibration lines} (ACS) at the center of the k-space (the low frequency region) and relaxed the requirement of the exact estimation of receiver coil sensitivities. The AUTO-SMASH formulation can be written as:
\begin{equation}
\begin{split}
    & \textbf{y}(k_1,k_2 + m\Delta k_2) =\\ &\sum_{i=1}^{n}w_{i}^{0}\textbf{y}_{i}^{ACS}(k_1,k_2 + m\Delta k_2) = \int\int dn_1dn_2\sum_{i=1}^{n}w_{i}^{0}S_{i}\textbf{x}e^{-jk_1n_1 - j(k_2 + m\Delta k_2)n_2}
\end{split}
\end{equation}
We note that the AUTO-SMASH paper theoretically proved that it can learn a kernel that is, $\sum_{i=1}^{n}w_{i}^{m}\textbf{y}_{i}(k_1,k_2) = \sum_{i=1}^{n}w_{i}^{m}\textbf{y}_{i}^{ACS}(k_1,k_2 + m\Delta k_2)$, that is, it can learn a linear shift invariant convolutional kernel to interpolate missing k-space lines from the knowledge of the fully sampled k-space lines of ACS region. 
The variable density AUTO-SMASH (VD-AUTO-SMASH) \citep{VDAUTOSMASH} further improved the reconstruction process by acquiring multiple ACS lines in the center of the k-space. 
The composite signal $\textbf{x}$ is estimated by adding each individual component $\textbf{y}_i$ by deriving linear weights $w^{m}_{i}$ and thereby estimating the missing k-space lines. 
%%%%%%%%%%%%%%%%%%%
%%%%%%%%%%%
%%%%
%%
%
The more popular, generalized auto calibrating partially parallel acquisitions (GRAPPA) \citep{bouman1993generalized} method uses this flavour of VD-AUTO-SMASH, i.e. the shift-invariant linear interpolation relationships in k-space, to learn the coefficients of a convolutional kernel from the ACS lines. 
The missing k-space are estimated as a linear combination of observed k-space points coming from all receiver coils. 
The weights of the convolution kernel are estimated as follows: a portion of the k-space lines in the ACS region are artificially masked to get a simulated set of acquired k-space points $\textbf{y}^{ACS_1}$ and missing k-space points $\textbf{y}^{ACS_2}$. 
Using the acquired k-space lines $\textbf{y}^{ACS_1}$, we can estimate the weights of the GRAPPA convolution kernel $K$ by minimizing the following cost function:
%%%%%%%%%%%%%%%%
%%%%%%%%%
%%%
%
\begin{equation}
    \hat{K} = \arg\min_{K} ||\textbf{y}^{ACS_2} - K\circledast \textbf{y}^{ACS_1}||^{2}_{2}
\end{equation}
%%%%%%%%%%%%%%%%
%%%%%%%%%
%%%
%
where $\circledast$ represents the convolution operation. The GRAPPA method has shown very good results for uniform undersampling, and is the method used in product sequences by Siemens and GE scanners. 
There are also recent methods \citep{xu2018improved, chang2012nonlinear} that show ways to learn non-linear weights of a GRAPPA kernel.
%%%%%%%%%%%%%%%%%
%%%%%%%%%%%
%%%%%%%%
%%%
%

The GRAPPA method regresses k-space lines from a learned kernel without assuming any specific image reconstruction constraint such as sparsity, limited support, or smooth phase as discussed in \citep{kim2019loraki}. 
On the other hand, low-rank based methods assume an association between the reconstructed image and the k-space structure, thus implying that the convolution-structured Hankel or Toeplitz matrices leveraged from the k-space measurements must show a distinct null-space vector association with the kernel. 
As a result, any low-rank recovery algorithm can be used for image reconstruction. 
The simultaneous autocalibrating and k-space estimation (SAKE) \citep{shin2014calibrationless} algorithm used the block Hankel form of the local neighborhood in k-space across all coils for image reconstruction. 
Instead of using correlations across multiple coils, the low-rank matrix modeling of local k-space neighborhoods (LORAKS) \citep{haldar2013low} utilized the image phase constraint and finite image support (in image space) to produce very good image reconstruction quality. 
The LORAKS method does not require any explicit calibration of k-space samples and can work well even if some of the constraints such as sparsity, limited support, and smooth phase are not strictly satisfied. 
The AC-LORAKS \citep{haldar2015autocalibrated} improved the performance of LORAKS by assuming access to the ACS measurements, i.e.:
\begin{equation}
    \textbf{y} = \arg\min_{\textbf{y}} ||\textbf{y} - \mathcal{P}(\mathcal{M}A\textbf{x})N||^{2}_{2}
\end{equation}
where $\mathcal{P}(\cdot)$ is a mapping function that transforms the k-space measurement to a structured low-rank matrix, and the matrix $N$ is the null space matrix. 
The mapping $\mathcal{P}(\cdot)$ basically takes care of the constraints such as sparsity, limited support, and smooth phase. In the PRUNO \citep{zhang2011parallel} method, the mapping $\mathcal{P}(\cdot)$ only imposes limited support and parallel imaging constraints. 
On the other hand, the number of nullspace vectors in $N$ is set to 1 in the SPIRiT method \citep{lustig2010spirit}. 
The ALOHA method \citep{lee2016acceleration} uses the weighted k-space along with transform-domain sparsity of the image. 
Different from them, the method of \citep{otazo2015low} uses a spatio-temporal regularization. 

%%%%%%%%%%%%%%%%%%%%%%
%%%%%%%%%%%%%%
%%%%%%%%
%%
%
\subsection{Image Space Rectification based Methods}
\label{sec_classic_image}
%%%%%%%%%%%%%%%%%%%%
%%%%%%%%%%%
%%%%
%
%\paragraph{The Preparatory Methods} 
%The Non-uniform Fast Fourier Transform (NUFFT) \citep{fessler2003nonuniform} is based on the min-max interpolation based estimation. 
%The image is first transferred to a Fourier domain using the weighted k-point Fast Fourier Transform (FFT) and then the image is estimated back using a linear estimation of the weights. 
These methods directly estimate the image from k-space by imposing prior knowledge about the properties of the image (e.g., spatial smoothness). Leveraging image prior through linear interpolation works well in practice but largely suffers from sub-optimal solutions and as a result the practical cone beam algorithm \citep{laurette1996cone} was introduced that improves image quality in such a scenario. 
The sensitivity encoding (SENSE) method \citep{pruessmann1999sense} is an image unfolding method that unfolds the periodic repetitions from the knowledge of the coil. 
In SENSE, the signal in a pixel location $(i,j)$ is a weighted sum of coil sensitivities, i.e.;
\begin{equation}
    I_{k}(i,j) = \sum_{k=1}^{N_c}\sum_{j=1}^{N_2} S_{kj}\textbf{x}(i,j),
    \label{eq_100000}
\end{equation}
where $N_{2}$ is the height of image $\textbf{x}\in R^{N_{1}\times N_{2}}$, $N_c$ is the number of coils, and $ S_{k}$ is the coil sensitivity of the $k^{th}$ coil. 
The $I_{k}$ is the k$^{th}$ coil image that has aliased pixels at a certain position, and $i$ is a particular row and $j=\{1, \cdots, N_2\}$ is the column index counting from the top of the image to the bottom. 
The $S$ is the sensitivity matrix that assembles the corresponding sensitivity values of the coils at the locations of the involved pixels in the full FOV image $\textbf{x}$. 
The coil images $I_{k}$, the sensitivity matrix $S$, and the image $\textbf{x}$ in Eqn. \ref{eq_100000} can be re-written as;
\begin{equation}
    I = S\textbf{x}.
\end{equation}
By knowing the complex sensitivities at the corresponding positions, we can compute the generalized inverse of the sensitivity matrix:
\begin{equation}
    \textbf{x} = (\hat{S}^{H}\hat{S})^{-1}\hat{S}^{H}\cdot I.
\end{equation}

Please note that, $I$ represents the complex coil image values at the chosen pixel and has length $N_{c}$. 
In k-t SENSE and k-t BLAST \citep{tsao2003k} the information about the spatio-temporal support is obtained from the training dataset that helps to reduce aliasing.

%%%%%%%%%%%%%%%%%%%%
%%%%%%%%%%%
%%%%
%
%\paragraph{Physics Based Methods}

The physics based methods allow statistical modeling instead of simple geometric modeling present in classical methods and reconstruct the MR images using the underlying physics of the imaging system \citep{roeloffs2016model, tran2016model, maier2019rapid, tran2013model,hilbert2018accelerated, sumpf2011model, ben2016accelerated,zimmermann2017accelerated, schneider2020free}.
%The physics based method is also called as model based image reconstruction (MBIR). 
These types of methods sometimes use very simplistic anatomical knowledge based priors \citep{chen1991sensor, gindi1993bayesian, cao1997using} or ``pixel neighborhood'' \citep{szeliski2010computer} information via a Markov Random Field based regularization \citep{saccostochastic, besag1986statistical}. 

A potential function based regularization takes the form $\mathcal{R}(\textbf{x})=\sum_{i=2}^{n}\psi(\textbf{x}_{i} - \textbf{x}_{i-1})$, where the function, $\psi(\cdot)$, could be a hyperbolic, Gaussian \citep{bouman1993generalized} or any edge-preserving function \citep{thibault2007three}. 
The Total Variation (TV) could also be thought of as one such potential function. 
The \citep{rasch2018dynamic} shows a variational approach for the reconstruction of subsampled dynamic MR data, which combines smooth, temporal regularization with spatial total variation regularization.

Different from Total Variation (TV) approaches, \citep{bostan2012reconstruction} proposed a stochastic modeling approach that is based on the solution of a stochastic differential equation (SDE) driven by non-Gaussian noise. 
Such stochastic modeling approaches promote the use of nonquadratic regularization functionals by tying them to some generative, continuous-domain signal model.

%%%%%%%%%%%%%%%%%%%%
%%%%%%%%%%%
%%%%
%
%\paragraph{Compressed Sensing Based Methods} 

The Compressed Sensing (CS) based methods impose sparsity in the image domain by modifying Eqn. \ref{eq_1_cs_mri} to the following:
\begin{equation}
    \hat{\textbf{x}} = \min_{\textbf{x}}\frac{1}{2}||\textbf{y} - A\textbf{x}||^{2}_{2} + \lambda||\Gamma \textbf{x}||_{1},
\label{eq_6_compressed_sensing}
\end{equation}
where $\Gamma$ is an operator that makes $\textbf{x}$ sparse. 
The $l_{1}$ norm is used to promote sparsity in the transform or image domain. 
The $l_{1}$ norm minimization can be pursued using a basis pursuit or greedy algorithm \citep{boyd2004convex}. 
However, use of non-convex \textit{quasi}-norms \citep{chartrand2007exact, zhao2008iterative, chartrand2008restricted, saab2008stable} show an increase in robustness to noise and image non-sparsity.
The structured sparsity theory  \citep{boyer2019compressed} shows that only $\mathcal{O}(M+M\log N)$ measurements are sufficient to reconstruct MR images when $M$-sparse data with size $N$ are given. 
The kt-SPARSE approach of \citep{lustig2006kt} uses a spatio-temporal regularization for high SNR reconstruction. 

Iterative sparsity based methods \citep{ravishankar2012learning, liu2015balanced, liu2016projected} assume that the image can be expressed as a linear combination of the columns (atoms) from a dictionary $\Gamma$ such that $\textbf{x}=\Gamma^{T} \textbf{h}$ and $\textbf{h}$ is the coefficient vector. Hence Eqn. \ref{eq_3_tikonov} becomes:
\begin{equation}
    \begin{split}
        \hat{\textbf{x}} = \min_{\textbf{x}}\frac{1}{2}||\textbf{y} - A\textbf{x}||^{2}_{2} + \lambda\mathcal{R}(\textbf{x}); \quad \mathcal{R}(\textbf{x}) = \min_{\textbf{h}}\frac{1}{2}||\textbf{x} - \Gamma^{T} \textbf{h}||^{2}_{2} + \alpha||\textbf{h}||_{1}.
    \end{split}
\end{equation}
The SOUP-DIL method \citep{bruckstein2009sparse} uses an exact block coordinate descent scheme for optimization while a few methods \citep{chen2008prior, lauzier2012prior, chen2008prior} assume to have a prior image $\textbf{x}_{0}$, i.e. $\mathcal{R}(\textbf{x}-\textbf{x}_{0})$, to optimize Eqn. \ref{eq_3_tikonov}. 
The method in \citep{caballero2014dictionary} optimally sparsifies the spatio-temporal data by training an overcomplete basis of atoms. 

The method in \citep{CSRevisit} shows a DL based approach to leverage wavelets for reconstruction.
%%%%%%%%%%%%%%%%%%%%
%%%%%%%%%%%
%%%%
%
%\paragraph{Transform Based Models}

The transform based methods are a generalization of the CS approach that assumes a sparse approximation of the image along with a regularization of the transform itself, i.e., $\Gamma \textbf{x} = \textbf{h} + \epsilon$, where $\textbf{h}$ is the sparse representation of $\textbf{x}$ and $\epsilon$ is the modeling error. 
The method in \citep{ravishankar2015efficient} proposed a regularization as follows:
\begin{equation}
    \mathcal{R}(x) = \min_{\Gamma, h}\frac{1}{2}||\textbf{x} - \Gamma^{T} \textbf{h}||^{2}_{2} + \alpha Q(\Gamma),
\end{equation}
where $Q(\Gamma) = -\log |det \Gamma| + 0.5||\Gamma||^{2}$ is the transform regularizer. 
In this context, the STROLLR method \citep{wen2018power} used a global and a local regularizer. 
%%%%%%%%%%%%%%%%%%%%
%%%%%%%%%%%
%%%%
%
%\paragraph{Unrolled Optimization Methods} 

In general, Eqn. \ref{eq_3_deep_learning} is a non-convex function and cannot be optimized directly with gradient descent update rules. The unrolled optimization algorithm procedure decouples the data consistency term and the regularization term by leveraging variable splitting in Eqn \ref{eq_3_tikonov} as follows:
\begin{equation}
    \min_{x,z}||A\textbf{x}-\textbf{y}||^{2}_{2} + \mu||\textbf{x}-\textbf{h}||^{2}_{2} + \mathcal{R}(\textbf{h}),
\label{eq_physics}
\end{equation}
where the regularization is decoupled using a quadratic penalty on $\textbf{x}$ and an auxiliary random variable $\textbf{z}$. Eqn \ref{eq_physics} is optimized via alternate minimization of
\begin{equation}
\begin{split}
    & \textbf{h}_{i} = \min_{h} \lambda ||\textbf{x}_{i-1} - \textbf{h}||^{2}_{2} + \mathcal{R}(\textbf{h})
\end{split}
\label{eq_self_reg}
\end{equation}
and the data consistency term:
\begin{equation}
\begin{split}
    & \textbf{x}_{i} = \min_{x} ||A\textbf{x}_{i-1} - \textbf{y}||^{2}_{2} + \lambda||\textbf{h}_{i} - \textbf{x}||^{2}_{2}
\end{split}
\label{eq_self_dc}
\end{equation}
where the $\textbf{h}_{i}, \textbf{h}_{i-1}$ are the intermediate variables at iteration $i$. 
The alternating direction method of multiplier networks (ADMM net) introduce a set of intermediate variables, $\textbf{h}_{1}, \textbf{h}_{2}, \cdots, \textbf{h}_{n}$, and eventually we have a set of dictionaries, $\Gamma_{1}, \Gamma_{2}, \cdots, \Gamma_{n}$, such that, $\textbf{h}_{i} = \Gamma_{i}\textbf{x}$, collectively promote sparsity. The basic ADMM net update \citep{yang2018admm} is as follows:
\begin{equation}
    \begin{split}
        & \arg\min_{x} \frac{1}{2}||A\textbf{x}-\textbf{y}||^{2}_{2} + \sum_{i}||\Gamma_{i}\textbf{x}+\beta_{i}-\textbf{z}_{i}||^{2}_{2}; \arg\min_{\textbf{z}_{i}} \sum_{i}[\lambda_{i}g(\textbf{h}_{i})||\Gamma_{i}\textbf{x}+\beta_{i}-\textbf{h}_{i}||^{2}_{2}]\\
        & \qquad \qquad \qquad \qquad \beta_{i} \leftarrow \beta_{i-1} + \alpha_{i}(\Gamma_{i}\textbf{x} - \textbf{h}_{i})
    \end{split}
\end{equation}
where $g(\cdot)$ can be any sparsity promoting operator and $\beta$ is called a multiplier. 
The iterative shrinkage thresholding algorithm (ISTA) solves this CS optimization problem as follows:
\begin{equation}
    \begin{split}
        \textbf{h}_{i+1} = \textbf{x}_{i} - \Phi^{H}(\Phi \textbf{x}_{i} - y); \quad \textbf{x}_{i+1} = \arg\min_{x} \frac{1}{2}||\textbf{x} - \textbf{h}_{i+1}||^{2}_{2} + \lambda||\Gamma \textbf{x}||_{1}.
    \end{split}
\end{equation}
Later in this paper, we shall show how ISTA and ADMM can be organically used within the modern DL techniques in Sec. \ref{sec_nongen_cnn}.
\section{Review of Deep Learning Building Blocks}
\label{sec_DLBB}
In this section, we will describe basic building blocks that are individually or collectively used  to develop complex DL methods that work in practice. 
Any DL method, by design, has three major components: the network structure, the training process, and the dataset on which the DL method is trained and tested. We shall discuss each one of them below in detail.
%%%%%%%%%%%%%%%%%%%
%%%%%%%%%%%%
%%%
%
\subsection{Various Deep Learning Frameworks}
\label{sec_DLBB_review}
%%%%%%%%%%%%%%%
%%%%%%%%%
%%%
%

{\bf Perceptron:} The journey of DL started in the year 1943 when Pitts and McCulloch \citep{mcculloch1943logical} gave the mathematical model of a biological neuron. This mathematical model is based on the ``all or none'' behavioral dogma of a biological neuron. Soon after, Rosenblatt provided the perceptron learning algorithm \citep{rosenblatt1957perceptron} which is a mathematical model based on the behaviour of a neuron. 
The perceptron resembles the structure of a neuron with dendrites, axons and a cell body. 
The basic perceptron is a binary classification algorithm of the following form:
%%%%%%%%%%%%%%5
\begin{equation}
    f(\mathbf{x}) = \begin{cases}1 & \text{if }\ \mathbf{w} \cdot \mathbf{x} + b > 0,\\0 & \text{otherwise}\end{cases}
\end{equation}
%%%%%%%%%%%%%%%%%%
where $\textbf{x}_{i}$'s are the components of an image vector $\textbf{x}$, $w_{i}$'s are the corresponding weights that determine the slope of the classification line, and $b$ is the bias term. This setup collectively resembles the ``all or none'' working principle of a neuron. However, in the famous book of Minsky and Papert \citep{minsky2017perceptrons} called ``Perpetron'' it was shown that the perceptron can't classify non-separable points such as an exclusive-OR (XOR) function.
%%%%%%%%%%%%%%%
%%%%%%%%%
%%%
%

\noindent {\bf Multilayer Perceptron (MLP):} It was understood that the non-separability problem of perceptron can be overcome by a multilayer perceptron \citep{minsky2017perceptrons} but the research stalled due to the unavailability of a proper training rule. 
In the year 1986, Rumelhart \textit{et al.} \citep{rumelhart1986learning} proposed the famous ``backpropagation algorithm'' that provided fresh air to the study of neural network. A Multilayer Perceptron (MLP) uses several layers of multiple perceptrons to perform nonlinear classification. 
A MLP is comprised of an input layer, an output layer, and several densely connected in-between layers called \textit{hidden layers}: 
\begin{equation}
    \begin{split}
        & \textbf{h}^{1} = \psi^{1}\Big(\sum W^{1} \textbf{x} + b^{1}\Big); \quad \textbf{h}^{i} = \psi^{i}\Big(\sum W^{i} \textbf{h}^{i-1} + b^{i}\Big), \text{where,} \quad i\in\{2, \cdots, n-1\}\\
        & \textbf{y} = \psi^{n}\Big(\sum W^{n} \textbf{h}^{n-1} + b^{n-1}\Big).
    \end{split}
\end{equation}
Along with the hidden layers and the input-output layer, the MLP learns features of the input dataset and uses them to perform classification. 
The dense connection among hidden layers, input and output layers often creates a major computational bottleneck when the input dimension is very high. 
%%%%%%%%%%%%%%%
%%%%%%%%%
%%%
%

\noindent{\bf Neocognitron or the Convolutional Neural Network (CNN):} The dense connection (also known as global connection) of a MLP was too flexible a model and prone to overfitting and sometimes had large computational overhead. 
To cope with this situation, a local sliding window based network with shared weights was proposed in early 1980s called neocognitron network \citep{fukushima1982neocognitron} and later popularized as Convolutional Neural Network (CNN) in the year 1998 \citep{lecun1989backpropagation}. 
Similar to the formulation of \citep{liang2020deep}, we write the feedforward process of a CNN as follows:
\begin{equation}
    \begin{split}
        & C_{0} = \textbf{x}\\
        & C_{i} = \psi_{i-1}(K_{i-1}\star C_{i-1}); \quad \quad i\in\{1, \cdots, n-1\},\; \quad C_{n} = \psi_{n-1}(K_{n-1}\star C_{n-1}),
    \end{split}
    \label{eq_CNN}
\end{equation}
where $C_{i}\in R^{h\times w\times d}$ is the $i^{th}$ hidden layer comprised of $d$-number of feature maps each of size $h\times w$, $K_{i}$ is the $i^{th}$ kernel that performs a convolution operation on $C_{i}$, and $\psi_i(\cdot)$ are activation functions to promote non-linearity. 
We show a vanilla kernel operation in Eqn. \ref{eq_CNN}. 
Please note that, the layers of the CNN can either be a fully connected dense layer, a max-pooling layer that downsizes the input, or a dropout layer to perform regularization that is not shown in Eqn. \ref{eq_CNN}.
%%%%%%%%%%%%%%%
%%%%%%%%%
%%%
%

\noindent{\bf Recurrent Neural Networks (RNNs):} A CNN can learn hidden features of a dataset using its inherent deep structure and local connectivities through a convolution kernel. But they are not capable of learning the time dependence in signals. The recurrent neural network (RNN) \citep{rumelhart1986learning} at its basic form is a time series neural network with the following form:
\begin{equation}
    \begin{split}
        & \textbf{h}_{0} = \textbf{x} \\
        & \textbf{h}_{t} = \psi_{t}(W_{\textbf{hh}}\textbf{h}_{t-1} + W_{\textbf{xh}}\textbf{x}_{t}), \quad \quad t\in\{1, \cdots, n-1\}; \quad \textbf{h}_{n} = \psi_{n}(W_{\textbf{hy}}\textbf{h}_{n-1}),
    \end{split}
\end{equation}
where $t$ is the time and the RNN takes the input $\textbf{x}$ in a sequential manner. 
However, the RNN suffers from the problem of ``vanishing gradient''. 
The vanishing gradient is observed when gradients from output layer of a RNN trained with gradient based optimization method changes parameter values by a very small amount, thus effecting no change in parameter learning. 
The Long Short Term Memory (LSTM) network \citep{hochreiter1997long} uses memory gates, and sigmoid and/or tanh activation function and later ReLU activation function (see Sec. \ref{sec_DLBB_activation} for activation functions) to control the gradient signals and overcome the vanishing gradient problem. 
%%%%%%%%%%%%%%%%%%
%%%%%%%%%%%%
%%%%%
%

\noindent{\bf Transformer Networks:} Although the LSTM has seen tremendous success in DL and MR reconstruction, there are a few problems associated with the LSTM model \citep{vaswani2017attention} such as: (i) the LSTM networks perform a sequential processing of input; and (ii) the short attention span of the hidden states that may not learn a good contextual representation of input. Such shortcomings are largely mitigated using a recent advancement called the Transformer Network \citep{vaswani2017attention}. A transformer network has a self-attention mechanism\footnote{\textbf{Self-attention}: The attention mechanism provides a way to know which part of the input is to be given more focus. 
The self-attention, on the other hand, measures the contextual relation of an input by allowing it to interact with itself. 
Let's assume that we are at $\textbf{h}_i^{th}\in R^{c\times N}$ layer that has $C$ number of channels and $N$ number of locations on a feature map. We get two feature vectors, $f(\textbf{h}_{i}) = W_{f}\textbf{h}_{i}$ and $g(\textbf{h}_{i}) = W_{g}\textbf{h}_{i}$, by transforming the $\textbf{h}_i^{th}$ layer to a vector (typically done with a $1\times 1$ convolution). 
The contextual similarity of these to vectors $f(\textbf{h}_{i})$ and $g(\textbf{h}_{i})$ is measured by;
\begin{equation} \nonumber
    \beta_{k,l} = \frac{\exp(\textbf{s}_{kl})}{\sum_{l=1}^{N}\exp(\textbf{s}_{kl})}, \text{where,} \textbf{s}_{kl} = f(\textbf{h}_{i})^{T} g(\textbf{h}_{i})
\end{equation}
The output is $\textbf{o} = (\textbf{o}_1, \textbf{o}_2, \cdots, \textbf{o}_N)$, where $\textbf{o}_{k} = v(\sum_{l=1}^{N}\beta_{k,l}m(\textbf{h}_{il}))$, and $m(h_{il}) = W_{m}h_{i}$, $v(h_{il}) = W_{v}\textbf{h}_{il}$. 
Here, the $W_{v}, W{m}, W_{f}, W_{g}$ are learnable matrices that collectively provide the self-attention vector $o\textbf{o} = (\textbf{o}_1, \textbf{o}_2, \cdots, \textbf{o}_N)$ for a given layer $\textbf{h}_{i}$. 
}, a positional embedding and a non-sequential input processing setup and empirically this configuration outperforms the LSTM networks by a large margin \citep{vaswani2017attention}.
%%%%%%%%%%%%%%
%%%%%%%%
%%
%
\subsection{Activation Functions}
\label{sec_DLBB_activation}
The activation function, $\psi(\cdot)$, operates on a node or a layer of a neural network and provides a boolean output, probabilistic output or output within a range. The step activation function was proposed by McCulloch \& Pitts in \citep{mcculloch1943logical} with the following form: $\psi(\textbf{x}) = 1, \text{if}, \textbf{x}\geq0.5$ and 0 otherwise. Several initial works also used the hyperbolic tangent function, $\psi(\textbf{x}) = \tanh(\textbf{x})$, as an activation function that provides value within the range $[-1, +1]$. The sigmoid activation function, $\psi(\textbf{x}) = \frac{1}{1+e^{-x}}$, is a very common choice and provides only positive values within the range $[0,1]$. However, one major disadvantage of the sigmoid activation function is that its derivative, $\psi^{'}(\textbf{x}) = \psi(\textbf{x})(1-\psi(\textbf{x}))$, quickly saturates to zero which leads to the \textit{vanishing gradient} problem. This problem was addressed by adding a Rectified Linear Unit (ReLu) to the network \citep{brownlee2019gentle}, $\psi(\textbf{x}) = \max(0, \textbf{x})$ with the derivative $\psi^{'}(\textbf{x}) = 1, if, \textbf{x}>0$ or 0 elsewhere.
%%%%%%%%%%%%%%%%%%%%%%
%%%%%%%%%%%%%%%
%%%%%%%%%
%%
%

\subsection{Network Structures}
\label{sec_DLBB_networks}
%%%%%%%%%%%
%%%%%%%
%%%
%

\noindent{\bf The VGG Network:} In late 2015, Zisserman \textit{et al.} published their seminal paper with the title ``very deep convolutional networks for large-scale image recognition'' (VGG) \citep{simonyan2014very} that presents a 16-layered network called VGG network. Each layer of the VGG network has an increasing number of channels. The network was shown to achieve state-of-the-art level performance in Computer Vision tasks such as classification, recognition, etc.   
%%%%%%%%%%%%%%%%
%%%%%%
%%
%
%%%%%%%%%%%%%%%%%
%%%%%%%%%%
%%%%
%%
%
\begin{figure*}[!ht]
    \centering
    \includegraphics[width=\textwidth]{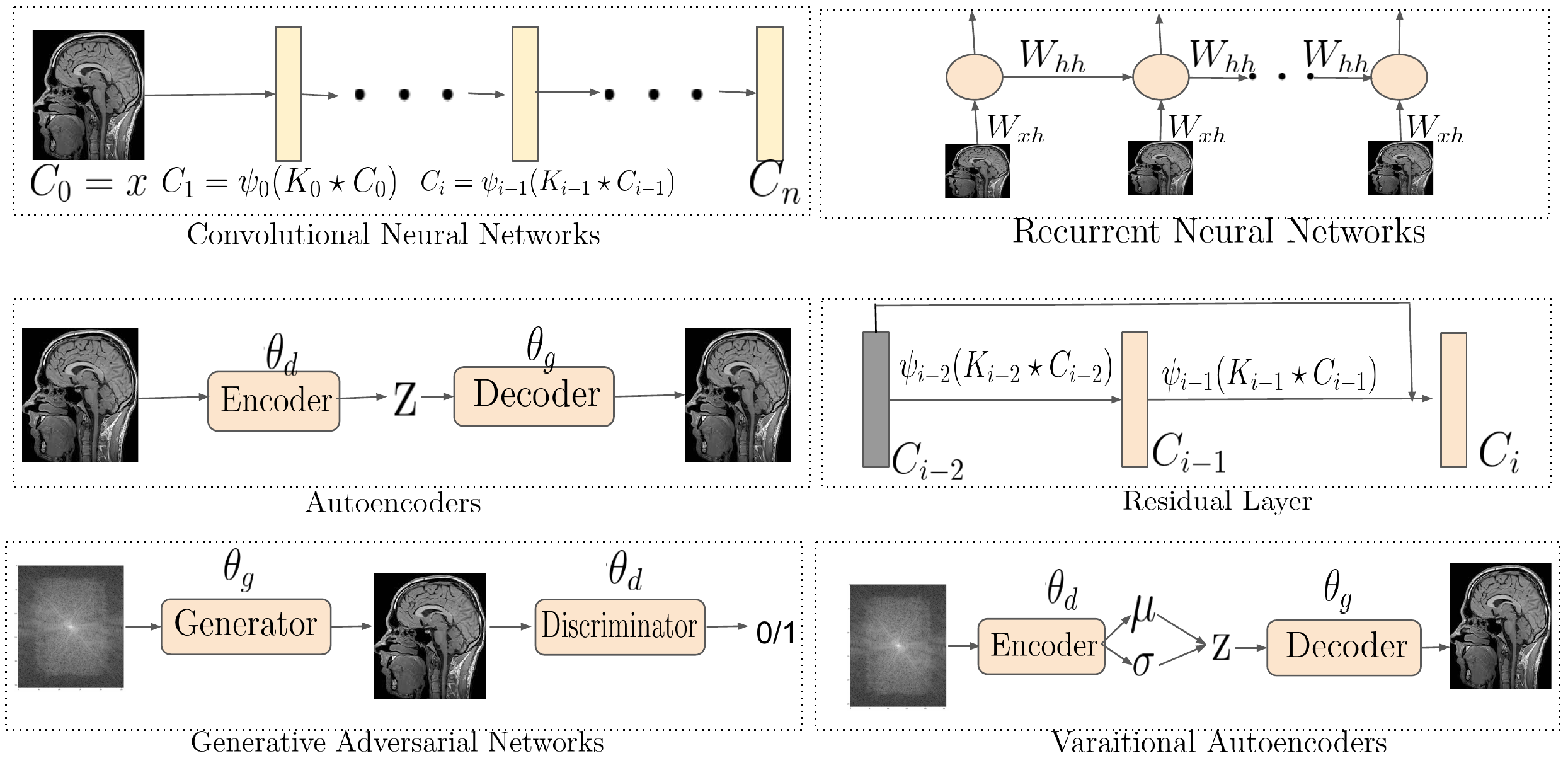}
    \caption{\textbf{Deep Learning Models}: We visually demonstrate various DL models such as Convolutional Neural Networks (CNNs), Recurrent Neural Networks (RNNs), Autoencoders, Residual Layer, Generative Adversarial Networks (GANs) and Variational Autoencoders (VAEs) discussed in Sec. \ref{sec_DLBB}. The \textbf{\textit{Convolutional Neural Network}} is comprised with many layers $\{C_{0}, C_{1}, \cdots, C_{n}\}$ and the network gets input for layer $C_{i+1}$ from its previous layer $C_{i}$ after imposing non-linearity, i.e. $C_{i+1} = \psi_{i}(K_{i}\star C_{i})$, using a non-linear function $\psi_{i}$. The \textit{\textbf{Recurrent Neural Network}} is a time varying memory network. On the other hand, the \textit{\textbf{Autoencoders}} perform non-linear dimensionality reduction by projecting the input $\textbf{x}$ to a lower dimensional variable $\textbf{z}$ using an encoder network and projects back $\textbf{z}$ to the image space $\textbf{x}$ using a decoder network. The ResNet uses the \textit{\textbf{Residual Layer}} to solve problems like vanishing gradient, slow convergence, etc. The \textit{\textbf{Generative Adversarial Network}} and the \textit{\textbf{Variational Autoencoder}} are generative models that are discussed in Sec. \ref{sec_generative_model}.}
    \label{fig_generative_models}
\end{figure*}
%%%%%%%%%%%%%%%%%
%%%%%%%%%%
%%%%
%%

\noindent{\bf The ResNet Model:}
A recently developed model called Residual Networks or ResNet \citep{he2016deep}, modifies the layer interaction shown in Eqn. \ref{eq_CNN} to the following form, $C_{i} = \psi_{i-1}(K_{i-1}\star C_{i-1}) + C_{i-2} \quad \text{where,} \quad i\in\{2, \cdots, n-1\}$, and provides a ``shortcut connection'' to the hidden layers. The identity mapping using the shortcut connections has a large positive impact on the performance and stability of the networks. 
%%%%%%%%%%%%%%%
%%%%%%%%
%%%
%
%

\noindent{\bf UNet:}
A UNet architecture \citep{ronneberger2015u} was proposed to perform image segmentation task in biomedical images. The total end-to-end architecture pictorially resembles the english letter ``U'' and has a encoder module and a decoder module. Each encoder layer is comprised of unpadded convolution, a rectified linear unit (ReLU in Sec. \ref{sec_DLBB_activation}), and a pooling layer, which collectively downsample the image to some latent space. The decoder has the same number of layers as the encoder. Each decoder layer  upsamples the data from its previous layer until the input dimension is reached. This architecture has been shown to provide good quantitative results on several datasets. 
%%%%%%%%%%%%%%%
%%%%%%%%
%%%

\noindent{\bf Autoencoders:}
The autoencoders (AE) are a type of machine learning models that capture the patterns or regularities of input data samples in an unsupervised fashion by mapping the target values to be equal to the input values (i.e. identity mapping). For example, 
given a data point $\textbf{x}$ randomly sampled from the training data distribution $p_{data}(\textbf{x})$, a standard AE learns a low-dimensional representation $\textbf{z}$ using an encoder network, $\textbf{z}\sim D(\textbf{z}|\textbf{x}, \theta_{d})$, that is parameterized by $\theta_d$. 
The low-dimensional representation $\textbf{z}$, also called the latent representation, is subsequently projected back to the input dimension using a decoder network, $\tilde{\textbf{x}}\sim G_{GEN}(\textbf{x}|\textbf{z},\theta_{g})$, that is parameterized by $\theta_g$. The model parameters, i.e. $(\theta_d, \theta_g)$, are trained using the standard back propagation algorithm with the following optimization objective: 
\begin{equation}
\begin{split}
    \mathcal{L}_{\theta_{d}, \theta_g} = \arg\min_{\theta_g, \theta_d}\mathbb{E}_{\textbf{z}\sim D(\textbf{x},\theta_{d})}\frac{1}{2}||\textbf{x} - G_{GEN}(\textbf{x}|\textbf{z},\theta_{g})||^{2}_{2}.
\end{split}
\label{eq_autoencoder}
\end{equation}
From a Bayesian learning perspective, an AE learns a posterior density, $p(\textbf{z}|\textbf{x})$, using the encoder network, $G_{GEN}(\textbf{z}|\textbf{x},\theta_{g})$, and a decoder network, $D(\textbf{x}|\textbf{z},\theta_{d})$. 
The vanilla autoencoder, in a way, can be thought of as a non-linear principal component analysis (PCA) \citep{hinton2006reducing}
that progressively reduces the input dimension using the encoder network and finds regularities or patterns in the data samples. 
%%%%%%%%%%%%%%%
%%%%%%%%
%%%
%

\noindent{\bf Variational Autoencoder Networks:}
Variational Autoecoder (VAE) is basically an autoencoder network that is comprised of an encoder network $G_{GEN}(\textbf{z}|\textbf{x},\theta_{g})$ that estimates the posterior distribution $p(\textbf{z}|\textbf{x})$ and the inference $p(\textbf{x}|\textbf{z})$ with a decoder network $D(z,\theta_{d})$. However, the posterior $p(\textbf{z}|\textbf{x})$ is intractable\footnote{\textbf{Intractability}: We can learn the density of the latent representation from the data points themselves, i.e. $p(\textbf{z}|\textbf{x}) = p(\textbf{z})p(\textbf{x}|\textbf{z})|p(\textbf{x})$, by expanding the Bayes theorem of conditional distribution. In this equation, the numerator is computed for a single realization of data points. However, the denominator is the marginal distribution of data points, $p(\textbf{x})$, which are complex and hard to estimate; thus, leading to intractability of estimating the posterior.} 
and several methods have been proposed to approximate the inference using techniques such as the Metropolis-Hasting \citep{metro} and variational inference (VI) algorithms \citep{vae}. The main essence of VI algorithm is to estimate an intractable probability density, i.e. $p(\textbf{z}|\textbf{x})$ in our case, from a class of tractable probability densities, $\mathcal{Z}$, with an objective to finding a density, $q(\textbf{z})$, almost similar to $p(\textbf{z}|\textbf{x})$. We then sample from the tractable density $q(\textbf{z})$ instead of $p(\textbf{z}|\textbf{x})$ to get an approximate estimation, i.e.
\begin{equation}
    q^{*}(z) = \arg\min_{q(\textbf{z})\in\mathcal{Q}} D_{KL}(q(\textbf{z})||p(\textbf{z}|\textbf{x})).
\end{equation}
Here, $D_{KL}$ is the KL divergence \citep{kl}. The VI algorithm, however, typically never converges to the global optimal solution but provides a very fast approximation. The VAE consists of three components: (i) the encoder network $E_{\theta_e}(\textbf{z}|\textbf{x})$ that observes and encodes a data point $\textbf{x}$ from the training dataset $\mathcal{D}(\textbf{x})$ and provides the mean and variance of the approximate posterior, i.e. $\mathcal{N}(\mu_{\theta}(\textbf{x}_{i}), \sigma_{\theta}(\textbf{x}_{i}))$ from a batch of $n$ data points $\textbf{x}_{i}|_{i=1}^{n}$; (ii) a prior distribution $G(z)$, typically an isotropic Gaussian $\mathcal{N}(0,I)$ from which $\textbf{z}$ is sampled, and (iii) a generator network $G_{\theta_g}(\textbf{x}|\textbf{z})$ that generates data points given a sample from the latent space $\textbf{z}$. 
The VAE, however, cannot directly optimize the VI, i.e. $q^{*}(z) = \arg\min_{q(\textbf{z})\in\mathcal{Q}} D_{KL}(q(\textbf{z})||p(\textbf{z}|\textbf{x}))$, as the KL divergence requires an estimate of the intractable density $p(\textbf{x})$, i.e. $D_{KL}(q(\textbf{z})||p(\textbf{z}|\textbf{x})) = \mathbb{E}[\log q(\textbf{z})] - \mathbb{E}[\log p(\textbf{z}, \textbf{x})] + \mathbb{E}[\log p(\textbf{x})]$. As a result, VAE estimates the evidence lower bound (ELBO) that is similar to KL divergence, i.e.:
\begin{equation}
\label{eq_elbo}
\begin{split}
    & ELBO(q) = \mathbb{E}[\log p(\textbf{z}, \textbf{x})] - \mathbb{E}[\log q(\textbf{z})] = \mathbb{E}[\log p(\textbf{z})] + \mathbb{E}[\log p(\textbf{x}|\textbf{z})] - \mathbb{E}[\log q(\textbf{z})]\\
    & \qquad \qquad = \mathbb{E}[\log p(\textbf{x}|\textbf{z})] - D_{KL}(p(\textbf{z}|\textbf{x})||q(\textbf{z}))= \mathbb{E}_{\textbf{z}\sim p(\textbf{z}|\textbf{x})}[\log p(\textbf{x}|\textbf{z})] - D_{KL}(p(\textbf{z}|\textbf{x})||q(\textbf{z}))
\end{split}
\end{equation}
Since $ELBO(q) \leq \log p(\textbf{x})$, optimizing Eqn. \eqref{eq_elbo} provides a good approximation of the marginal density $p(\textbf{x})$. Please note that, Eqn. \ref{eq_elbo} is similar to Eqn. \ref{eq_3_tikonov} where the first term in Eqn. \ref{eq_elbo} is the data consistency term and the KL divergenece term acts as a regularizer.
%%%%%%%%%%%%%%%%
%%%%%%
%%
%

\noindent{\bf Generative Adversarial Networks:}
A vanilla Generative Adversarial Networks (GAN) setup, by design, is an interplay between two neural networks called the \textit{generator}, that is $G_{GEN}(\textbf{x}|\textbf{z},\theta_{g})$, and the \textit{discriminator} $D_{\theta_d}(\cdot)$ parameterized by $\theta_{g}$ and $\theta_{d}$ respectively. The $G_{GEN}(\textbf{x}|\textbf{z},\theta_{g})$ samples the latent vector $z\in\mathbb{R}^{n\times1\times1}$ and generates \generated. While the discriminator, on the other hand, takes $\textbf{x}$ (or \generated) as input and provides a $\{real,fake\}$ decision on $\textbf{x}$ being sampled from a real data distribution or from $G_{GEN}(\textbf{x}|\textbf{z},\theta_{g})$. The parameters are trained using a game-theoretic adversarial objective, i.e.:
\begin{equation}
\begin{split}
    &\quad \mathcal{L}_{\theta_{d}, \theta_g} = -\mathbb{E}_{\textbf{x}\sim p_{data}}[\log D(\textbf{x}|\theta_{d})]  - \mathbb{E}_{\textbf{z}\sim \mathcal{N}(0,I)}[\log(1-D(G_{GEN}(\textbf{x}|\textbf{z},\theta_{g}),\theta_{d}))].
\end{split}
\label{eq_GAN}
\end{equation}
As the training progresses, the generator $G_{GEN}(\textbf{z}|\theta_{g})$ progressively learns a strategy to generate realistic looking images, while the discriminator $D(\textbf{x},\theta_{d})$ learns to discriminate the generated and real samples. %In this manner, a GAN can capture the conditional distribution $p(\textbf{x}|c)$, where $c$ is a random variable the distribution is conditioned on, and the joint distribution, $G(\textbf{x}_{1}, x_{2}, \cdots, x_{m})$, where $\{x_{1}, x_{2}, \cdots, x_{k}\}$ is a set of $k$ random variables.

%%%%%%%%%%%%%%%
%%%%%%%%
%%%
%

\subsection{Loss Functions}
\label{sec_DLBB_loss} 
In Sec. \ref{sec_intro_mathematical_formulaion}, we mentioned the loss function, $L$, that estimates the empirical loss and the generalization error. 
 Loss functions that are typically used in MR image reconstruction using DL methods are the Mean Squared Error (MSE), Peak Signal to Noise Ratio (PSNR) and the Structural Similarity Loss function (SSIM), or an $l_{1}$ loss to optimize Eqns. \ref{eq_2_least_square}, \ref{eq_3_tikonov},  \ref{eq_3_deep_learning}. 
The MSE loss between an image $\textbf{x}$ and its noisy approximation $\hat{\textbf{x}}$ is defined as, $MSE(\textbf{x}, \hat{\textbf{x}}) = \frac{1}{n}\sum_{i=1}^{n}(\hat{\textbf{x}}_{i}-\textbf{x}_{i})^{2}$, where $n$ is the number of samples. 
The root MSE (RMSE) is essentially the squared root of the MSE, $\sqrt{MSE(\textbf{x}, \hat{\textbf{x}})}$. The $l_{1}$ loss, $l_{1}(\textbf{x}, \hat{\textbf{x}}) = \frac{1}{n}\sum_{i=1}^{n}|\hat{\textbf{x}}_{i}-\textbf{x}_{i}|$, provides the absolute difference and is typically used as a regularization term to promote sparsity. 
The PSNR is defined using the MSE, PSNR$(\textbf{x}, \hat{\textbf{x}}) = 20\log_{10}(MAX_{\textbf{x}} - 10\log_{10}(MSE(\textbf{x}, \hat{\textbf{x}}))$, where $MAX_{\textbf{x}}$ is the highest pixel value attained by the image $\textbf{x}$. 
The PSNR metric captures how strongly the noise in the data affects the fidelity of the approximation with respect to the maximum possible strength of the signal (hence the name peak signal to noise ratio). 
The main concern with $MSE(\textbf{x}, \hat{\textbf{x}})$ and PSNR$(\textbf{x}, \hat{\textbf{x}})$ is that they penalize large deviations much more than smaller ones \citep{zhao2016loss} (e.g., outliers are penalized more than smaller anatomical details).
The SSIM loss for a pixel $i$ in the image $\textbf{x}$ and the approximation $\hat{\textbf{x}}$ captures the perceptual similarity of two images: SSIM$(\textbf{x}_{i}, \hat{\textbf{x}}_{i}) = \frac{2\mu_{\textbf{x}}\mu_{\hat{\textbf{x}}}+c_{1}}{\mu_{\textbf{x}}^{2}+\mu_{\hat{\textbf{x}}}^{2}+c_{1}}\frac{2\sigma_{\textbf{x}\hat{\textbf{x}}}+c_{2}}{\sigma_{\textbf{x}}^{2}+\sigma{\hat{\textbf{x}}}^{2}+c_{2}}$, here $c_1, c_2$ are two constants, and $\mu$ is the mean and $\sigma$ is the standard deviation. 

\noindent{\bf The VGG loss:} It is shown in \citep{johnson2016perceptual} that the deeper layer feature maps (feature maps are discussed in Eqn. \ref{eq_CNN}) of a VGG-16 network, i.e. a VGG network that has 16 layers, can be used to compare perceptual similarity of images. 
Let us assume that, the $L^{th}$ layer of a VGG network has distinct $N_{L}$ feature maps each of size $M_{L} \times M_{L}$. 
The matrix $F^{L}\in \mathbb{R}^{N_{L}\times M_{L}}$, stores the activations $F^{L}_{i,j}$ of the $i^{th}$ filter at position $j$ of layer $L$. 
Then, the method computes feature correlation using: $C^{L}_{i,j} = \sum_{k} F_{i,k}^{L} F_{j,k}^{L}$, where any $F^{m}_{n,o}$ conveys the activation of the $n^{th}$ filter at position $o$ in layer $m$. 
The correlation $C^{L}_{i,j}$ is considered as a VGG-loss function. 

%%%%%%%%%%%%%
%%%%%%
%%
%

\section{Inverse Mapping using Deep Generative Models}
\label{sec_generative_model}
Based on how the generator network, $G_{GEN}(\textbf{x}|\textbf{z},\theta_{g})$, is optimized in Eqn. \ref{eq_4_generative_models}, we get different manifestations of deep generative networks such as Generative Adversarial Networks, Bayesian Networks, etc. In this section, we shall discuss specifics on how these networks are used in MR reconstruction.
%%%%%%%%%%%
%%%%%%%
%%%
%
\subsection{Generative Adversarial Networks (GANs)}
\label{sec_GAN}
%%%%%%%%%%%%%%
%%%%%
%
We gave a brief introduction to GAN in Sec. \ref{sec_DLBB_networks} and in this section we shall take a closer look on the different GAN methods used to learn the inverse mapping from k-space measurements to the MR image space.
%%%%%%%%%%%%%%%
%%%%%%%%%%
%%%%%%
%%
%

\noindent{\bf Inverse Mapping from k-space:}
The current GAN based k-space methods can be broadly classified into two categories: (i) methods that directly operate on the k-space $\textbf{y}$ and reconstruct the image $\textbf{x}$ by learning a non-linear mapping and (ii) methods that impute the missing k-space lines $\textbf{y}^{missing}$ in the undersampled k-space measurements. 
In the following paragraphs, we first discuss the recent GAN based direct k-space to MR image generation methods followed by the undersampled k-space to full k-space generation methods. 

%Correction of motion artefact has been a more investigated topic in direct k-space to MR image generation. 
%The GAN-based methods have offered good improvements in this direction. 
The direct k-space to image space reconstruction methods, as shown in Fig. \ref{fig_generative_models} (a), are based on the premise that the missing k-space lines can be estimated from the acquired k-space lines provided we have a good non-linear interpolation function, i.e.;
\begin{equation}
\begin{split}
    &\quad \mathcal{L}_{\theta_{d}, \theta_g} = -\mathbb{E}_{\textbf{x}\sim p_{data}}[\log D(\textbf{x}|\theta_{d})]  - \mathbb{E}_{z\sim \mathcal{N}(0,I)}[\log(1-D(G_{GEN}(\textbf{x}|y,\theta_{g}),\theta_{d}))].
\end{split}
\label{eq_direct_k_image}
\end{equation}
This GAN framework was used in \citep{oksuz2018cardiac} for correcting motion artifacts in cardiac imaging using a generic AUTOMAP network\footnote{\textbf{AUTOMAP} \citep{Automap}: This is a two stage network resembling the unrolled optimization like methods \citep{schlemper2017deep}. The first sub network ensures the data consistency, while, the other sub network helps in refinement of the image. The flexibility of AUTOMAP enables it to learn the k-space to image space mapping from alternate domains instead of strictly from a paired k-space to MR image training dataset.}. 
%On UK Biobank dataset \citep{UKBIO} consisting of
%2000 good quality CINE MR images the method has shown $0.850$ SSIM and $35.1$ PSNR score. 
Such AUTOMAP-like generator architectures not only improve the reconstruction quality but help in other downstream tasks such as MR image segmentation \citep{AutomapGAN,oksuz2020deep, oksuz2019detection}. 
However, while the ``AUTOMAP as generator'' based methods solve the broader problem of motion artifacts, but they largely fail to solve the banding artifacts along the phase encoding direction. 
To address this problem, a method called MRI Banding Removal via Adversarial Training \citep{Banding} leverages a perceptual loss along with the discriminator loss in Eqn. \ref{eq_GAN}. 
The perceptual loss ensures data consistency, while, the discriminator loss checks whether:   
(i) the generated image has a horizontal ($0$) or a vertical ($1$) banding; and 
(ii) the generated image resembles the real image or not. 
With a 4x acceleration, a 12-layered UNet generator and a ResNet discriminator, the methodology has shown remarkable improvements \citep{Banding} on fastMRI dataset.

\begin{figure}[!t]
    \centering
    \includegraphics[width=0.8\textwidth]{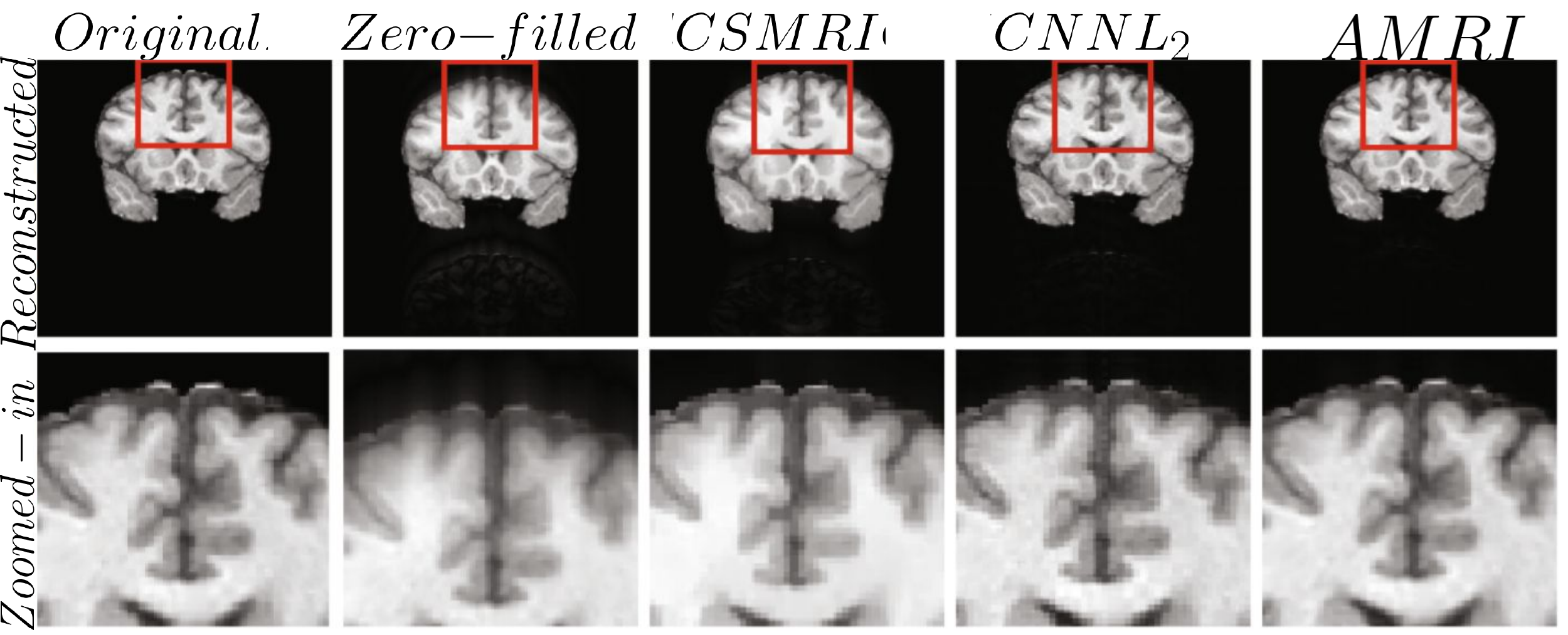}
    \caption{\textbf{Reconstruction using k-space Imputation}: We show qualitative comparison of a few methods with the AMRI method \citep{AMRI}. The first row are the reconstructed images and the second row images are zoomed-in version of the red boxes shown in first row. The Zero-filled, compressed sensing MRI (CS-MRI), CNN network trained with $l_2$ loss (CNN-L2) have lesser detail as compared to the AMRI (proposed) method. The images are borrowed from AMRI \citep{AMRI} paper. }
    \label{fig_AMRI}
\end{figure}

\begin{figure}[!t]
    \centering
    \includegraphics[width=0.6\textwidth]{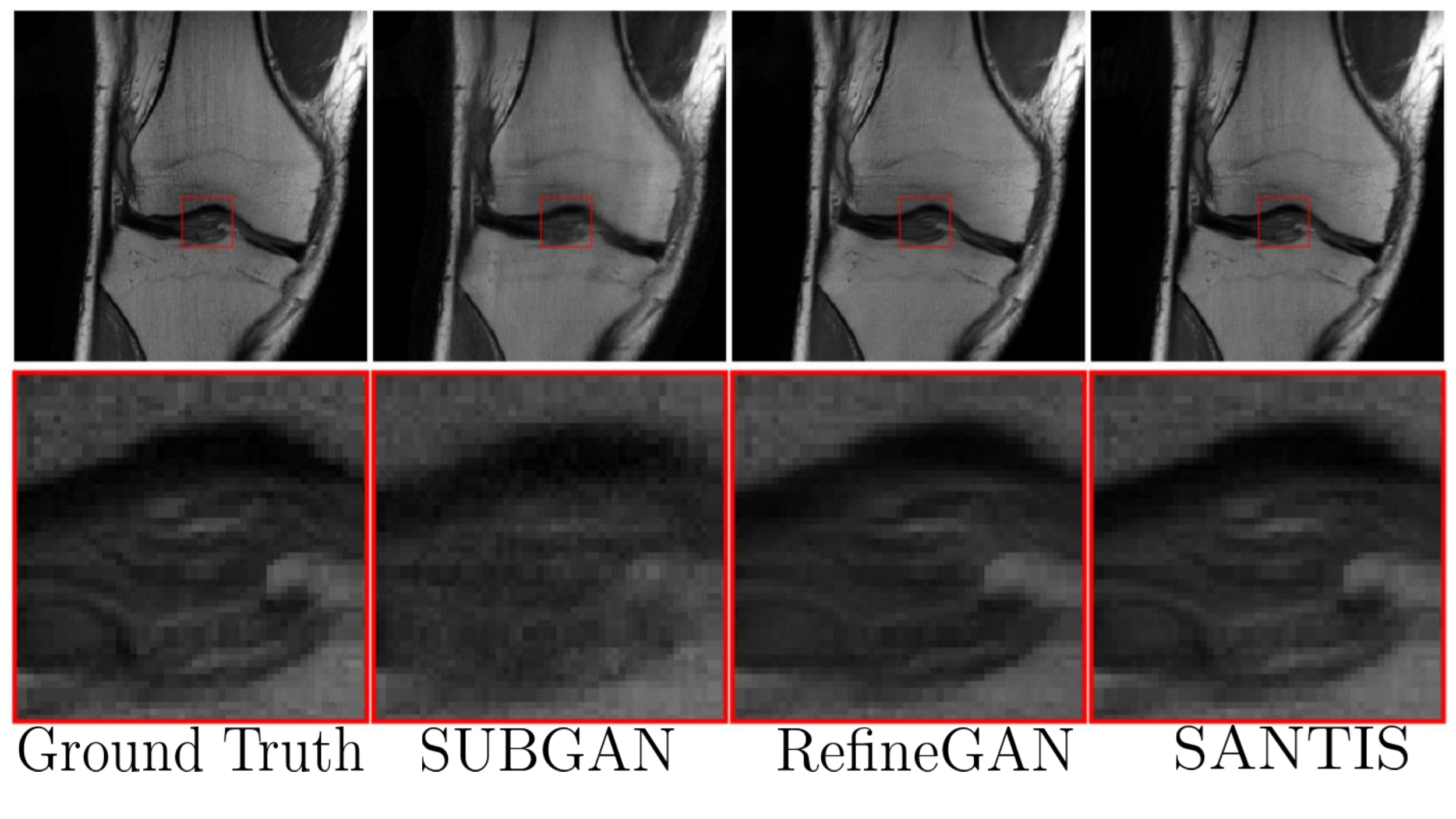}
    \caption{\textbf{Comparison of GAN Methods}: Qualitative comparison of (i) k-space interpolation based method, i.e. SUBGAN; (ii) Image space rectification method, i.e. RefineGAN; and (iii) the combined k-space and image space rectification method, i.e. SANTIS.}
    \label{fig_GAN_compare}
\end{figure}

Instead of leveraging the k-space regularization within the parameter space of a GAN \citep{oksuz2018cardiac, AutomapGAN}, the k-space data imputation using GAN directly operates on the k-space measurements to regularize Eqn. \ref{eq_GAN}. 
To elaborate, these type of methods estimate the missing k-space lines by learning a non-linear interpolation function (similar to GRAPPA) within an adversarial learning framework, i.e.
\begin{equation}
\begin{split}
    & \mathcal{L}_{\theta_{d}, \theta_g} = -\mathbb{E}_{\textbf{x}\sim p_{data}}[\log D(\textbf{x}|\theta_{d})] - \mathbb{E}_{\textbf{z}\sim \mathcal{N}(0,I)}[\log(1-D(\mathscr{F}G_{GEN}(\textbf{y}^{full}|\textbf{y},\theta_{g}),\theta_{d}))].
\end{split}
\label{eq_k_k}
\end{equation}
The accelerated magnetic resonance imaging (AMRI) by adversarial neural network method \citep{AMRI} aims to generate the missing k-space lines, $\textbf{y}^{missing}$ from $y$ using a conditional GAN, $\textbf{y}^{missing} \sim G(\textbf{y}^{missing}|\textbf{z},c=y)$. 
The combined $y, \textbf{y}^{missing}$ is Fourier transformed and passed to the discriminator. 
The AMRI method showed improved PSNR value with good reconstruction quality and no significant artifacts as shown in Fig. \ref{fig_AMRI}.
%%%%%%%%%%%%%%%%%%
%%%%%%%%%%%
%%%%%
%%
%
%%%%%%%%%%%%%%%%%
%%%%%%%%%%
%%%%
%%%
%%
%
Later, in subsampled brain MRI reconstruction by generative adversarial neural networks method (SUBGAN) \citep{SUBGAN}, the authors discussed the importance of temporal context and how that mitigates the noise associated with the target's movement. 
The UNet-based generator in SUBGAN takes three adjacent subsampled k-space slices $\textbf{y}_{i-1}, \textbf{y}_{i}, \textbf{y}_{i+1}$ taken at timestamps $t_{i-1}, t_{i}, t_{i+1}$ and provides the reconstructed image. 
The method achieved a performance boost of $\sim 2.5$ in PSNR with respect to the other state-of-the-art GAN methods while considering $20\%$ of the original k-space samples on IXI dataset \citep{rowland2004information}.  
We also show reconstruction quality of SUBGAN on fastMRI dataset in Fig. \ref{fig_GAN_compare}.
Another method called multi-channel GAN \citep{MULTI} advocates the use raw of k-space measurements from all coils and has shown good k-space reconstruction and lower background noise compared to classical parallel imaging methods like GRAPPA and SPIRiT. However, we note that this method achieved $\sim 2.8$ dB lower PSNR than the GRAPPA and SPIRiT methods.

Despite their success in MR reconstruction from feasible sampling patterns of k-space, the previous models we have discussed so far have the following limitations: 
(i) they need unaliased images for training, 
(ii) they need paired k-space and image space data, or
(iii) the need  fully sampled k-space data.
In contrast, we note a recent work called unsupervised MR reconstruction with GANs \citep{UnsupGAN} that only requires the undersampled k-space data coming from the receiver coils and optimizes a network for image reconstruction. 
Different from AutomapGAN \citep{AutomapGAN}, in this setup the generator provides the undersampled k-space (instead of the MR image as in case of AutomapGAN) after applying Fourier transform, sensitivity encoding and a random-sampling mask $\mathcal{M}_{1}$ on the generated image, i.e. $\textbf{y}^{gen} = \mathcal{M}_{1}\mathscr{F}\mathcal{S}(G(\textbf{x}|y, \theta_g))$. 
The discriminator takes the k-space measurements instead of an MR image and provides thee learned signal to the generator.
%%%%%%%%%%%
%%%%%%
%

\noindent{\bf Image space Rectification Methods:}
%K-space rectification based methods perform regularization in Eqn. \ref{eq_3_tikonov}, either approximating missing k-space line using the parameters of GAN, or learning a non-linear generator to estimate missing k-space lines. 
Image space rectification methods operate on the image space and learn to reduce noise and/or aliasing artifacts by updating Eqn. \ref{eq_3_tikonov} to the following form:
%%%%%%%%%%%%%%%%%
%%%%%%%%%%%%
%%%%%%
%%
%
\begin{equation}
\begin{split}
    & \hat{\textbf{x}} = \arg\min_{\textbf{x}}\frac{1}{2}||\textbf{x} - G_{GEN}(\textbf{x}|\textbf{x}^{low}, \theta_g)||^{2}_{2} + \lambda ||\textbf{y}^{t} - \mathscr{F}G_{GEN}(\textbf{x}|\textbf{x}^{low}, \theta_g)||^{2}_{2} + \zeta\mathcal{L}_{\theta_{d}, \theta_g};\\
    &\mathcal{L}_{\theta_{d}, \theta_g} = -\mathbb{E}_{\textbf{x}\sim p_{data}}[\log D(\textbf{x}|\theta_{d})] - \mathbb{E}_{z\sim \mathcal{N}(0,I)}[\log(1-D(G_{GEN}(\textbf{x}|\textbf{x}^{low},\theta_{g}),\theta_{d}))].
\end{split}
\label{eq_image_space_GAN}
\end{equation}
%%%%%%%%%%%%%%%%%
%%%%%%%%%%%%
%%%%%%
%%
%
The GAN framework for deep de-aliasing \citep{DAGAN} regularizes the reconstruction by adopting several image priors such as:
(i) image content information like object boundary, shape, and orientation, along with using a perceptual loss function: $\frac{1}{2}||\textbf{x} -\textbf{x}^{low}||_2^{2}$,  
(ii)  data consistency is ensured using a frequency domain loss $\frac{1}{2}||\textbf{y}^t - \textbf{y}^u||_2^{2}$, and 
(iii) a VGG loss (see Sec. \ref{sec_DLBB_loss}) which enforces semantic similarity between the reconstructed and the ground truth images. 
The method demonstrated a $2$dB improvement in PSNR score on IXI dataset with 30\% undersampling. 
However, it was observed that the finer details were lost during the process of de-aliasing with a CNN based GAN network. 
In the paper called self-attention and relative average discriminator based GAN (SARAGAN) \citep{SARAGAN}, the authors show that fine details tend to fade away due to smaller size of the convolution kernels, leading to poor performance. 
Consequently, the SARAGAN method adopts a relativistic discriminator \citep{Relativistic} along with a self-attention network (see Sec. \ref{sec_DLBB_review} for self-attention) to optimize the following equation that is different from Eqn. \ref{eq_image_space_GAN}:
%%%%%%%%%%%
%%%%%
%
\begin{equation}
\mathcal{L}_{\theta_{d}, \theta_g} = -\mathbb{E}_{\textbf{x}\sim p_{data}}[sigmoid(D(\textbf{x}) - D(G(\textbf{x}^{low})))] - \mathbb{E}_{z\sim \mathcal{N}(0,I)}[sigmoid(D(G(\textbf{x}^{low})) - D(\textbf{x}))]   
\end{equation}
%%%%%%%%%%%
%%%%%
%
where, $sigmoid(\cdot)$ is the sigmoid activation function discussed in Sec. \ref{sec_DLBB_activation}. 
This method showed excellent performance in the MICCAI 2013 grand challenge brain MRI reconstruction dataset and got an SSIM of $0.9951$ and PSNR of $45.7536 \pm 4.99$  with $30\%$ sampling rate. Among other methods, sparsity based constraints are imposed as a regularizer to Eqn. \ref{eq_image_space_GAN} in compressed sensing GAN (GANCS) \citep{mardani2018deep}, RefineGAN \citep{RefineGAN}, and the structure preserving GAN \citep{StructurePreservingGAN, lee2018deep} methods. 
Some qualitative results using the RefineGAN method are shown in Fig. \ref{fig_GAN_compare}. 
On the other hand, methods like PIC-GAN \citep{PICGAN} and (MGAN) \citep{MGAN} use a SENSE-like reconstruction strategy that combines MR images reconstructed from parallel receiver coils using a GAN framework. 
Such methods have also shown good performance with low normalized mean squared error in the knee dataset.
%%%%%%%%%%
%%%%
%

\noindent{\bf Combined k-space and image space methods:}
Thus far, we have discussed k-space (GRAPPA-like GAN methods) and image space (SENSE-like GAN methods) MR reconstruction methods that work in isolation.
However, both these strategies can be combined together to leverage the advantages of both methods.
Recently, a method called sampling augmented neural network with
incoherent structure for MR image reconstruction (SANTIS) \citep{SANTIS} was proposed that
leverages a cycle consistency loss, $\mathcal{L}_{cyc}$ in addition to a GAN loss, $\mathcal{L}_{\theta_d, \theta_g}$ i.e.,
\begin{equation}
    \begin{split}
        & \mathcal{L}_{full} = \mathcal{L}_{\theta_d, \theta_g} + \mathcal{L}_{cyc} = \lambda_{1}\mathbb{E}[||\textbf{x} - G(\textbf{x}^{low}, \theta_g)||_{2}^{2}] + \lambda_{GAN}\Big(\mathbb{E}[\log D(\textbf{x}, \theta_d)]\\
        &\qquad +\mathbb{E}[\log(1-D(G(\textbf{x}^{low}, \theta_g), \theta_d))]\Big)+ \lambda_{2}\mathbb{E}[||y - F(G(\textbf{x}^{low},\theta_{g}), \theta_{f})||_{2}^{2}],
    \end{split}
\end{equation}
where the function $F(\cdot)$ is another generator network that projects back the MR image to k-space. 
The method achieved an SSIM value of $91.96$ on the 4x undersampled knee FastMRI dataset (see Fig. \ref{fig_GAN_compare} and Table 1). 
In the collaborative GAN method (CollaGAN) \citep{COLLAGAN}, instead of cycle consistency between k-space and the image domain from a single image, they consider a collection of domains such as T1-weighted and T2-weighted data and try to reconstruct the MR images with cycle consistency in all domains. 
The InverseGAN \citep{InverseGAN} method performs cycle consistency using a single network that learns both the forward and inverse mapping from and to k-space. 
%%%%%%%%%%%%%
%%%%
%
%%%%%%%%%%%
%%%%
%%
%
\subsection{Bayesian Learning}
\label{sec_vaes}
\begin{figure}[!t]
    \centering
    \includegraphics[width=0.6\textwidth]{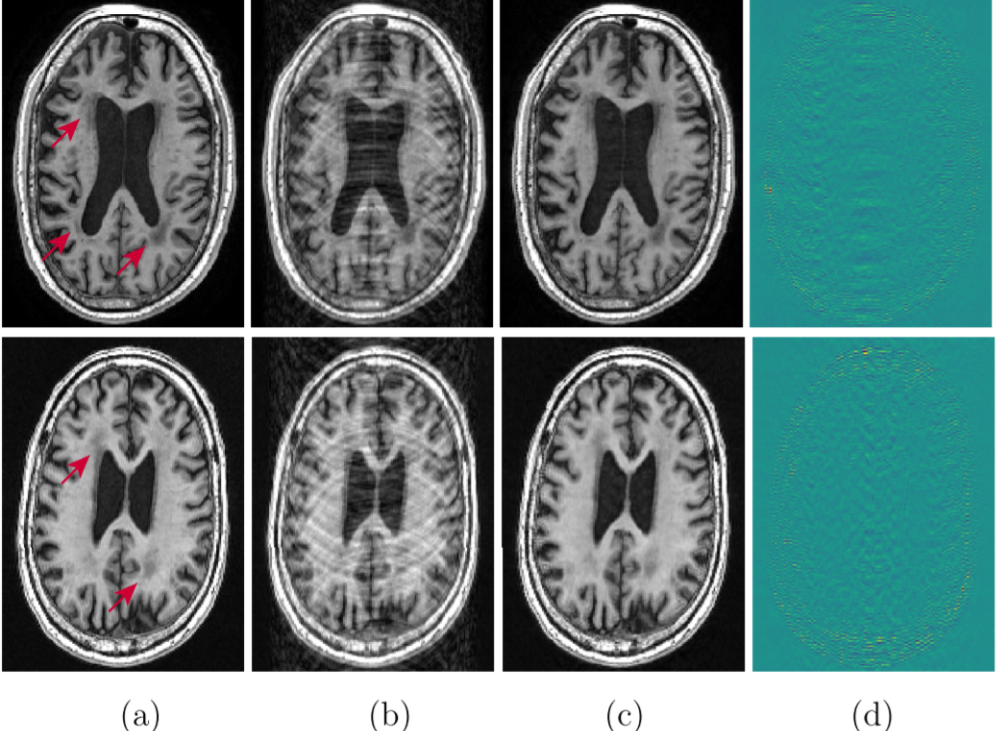}
    \caption{\textbf{MR Reconstruction using VAE Based Methods}: Qualitative comparison of (a) original; (b) zero-filled; and (c) VAE based DDP method \citep{DDP}. Also shown are the difference maps in (d) between DDP and the original images. These results are borrowed from the DDP \citep{DDP} paper.}
    \label{fig_DDP}
\end{figure}

Bayes’s theorem expresses the posterior $p(\textbf{x}|y)$ as a function of the k-space data likelihood, $p(y|\textbf{x})$ and the prior $p(\textbf{x})$ with the form $p(\textbf{x}|y)\propto p(y|\textbf{x})p(\textbf{x})$ also known as the ``product-of-the-experts'' in DL literature. 
In \citep{tezcan2018mr}, the prior is estimated with a Monte Carlo sampling technique which is computationally intensive. 
To overcome the computational cost, several authors have proposed to learn a non-linear mapping from undersampled k-space to image space using VAEs. 
In these VAE based methods \citep{DDP, JointDDP,van2012human}, the networks are trained on image patches $m_{j}$ obtained from k-space measurement $\textbf{y}_{i}$ and the VAE network is optimized using these patches with the following cost function: 
$\sum_{j=1}^{N} ELBO(m_j)$,
\begin{equation}
\begin{split}
    \arg\min_{m} \Big[||E\textbf{x}-y||^{2}_{2}-\sum_{\textbf{x}\in\Omega(m)}ELBO(\textbf{x})\Big]; ELBO(\textbf{x}) = \mathbb{E}_{D_{\theta_d}(z|\textbf{x})} \Big[\log G_{\theta_g}(\textbf{x}|\textbf{z})+\log\frac{p(z)}{G_{\theta_g}(\textbf{x}|\textbf{z})}\Big].
\end{split}
\end{equation}
These methods have mainly been evaluated on the Human Connectome Project
(HCP) \citep{van2012human} dataset and have shown good performance on 4x undersampled images (see Fig \ref{fig_DDP}).

Different from them, PixelCNN+ considers each pixel as random variable and estimates the joint distribution of pixels over an image $\textbf{\textbf{x}}$ as the product of conditional distribution, i.e. $p(\textbf{x}) = \prod_{i=1}^{n^{2}}p(\textbf{x}_{i}|\textbf{x}_1, \textbf{x}_2, \cdots, \textbf{x}_{i-1})$. 
The method proposed in \citep{luo2020mri} considers a generative regression model called PixelCNN+ \citep{oord2016conditional} to estimate the prior $p(\textbf{x})$. 
This method demonstrated very good performance, i.e. they achieved more than 3 dB PSNR improvement than the current state-of-the-art methods like GRAPPA, variational networks (see Sec. \ref{sec_nongen_cnn}) and SPIRIT algorithms. 
The Recurrent Inference Machines (RIM) for accelerated MRI reconstruction \citep{lonning2018recurrent} is an general inverse problem solver that performs a step-wise reassessments of the maximum a posteriori and infers the inverse transform of a forward model. 
Despite showing good results, the overall computational cost and running time is very high compared to GAN or VAE based methods.
%%%%%%%%%%%
%%%%
%%
%
% Pre-rebuttal
%%%%%%%%%%%
%%%%
%%
%
% \subsection{Variational Autoencoder Networks (VAEs)}
% \label{sec_vaes}
% Apart from GAN based methods, several authors have proposed to learn a non-linear mapping from undersampled k-space to image space using VAEs. 
% In these VAE based methods \citep{DDP, JointDDP,van2012human}, the networks are trained on image patches $m_{j}$ obtained from k-space measurement $\textbf{y}_{i}$ and the VAE network is optimized using these patches with the following cost function: 
% $\sum_{j=1}^{N} ELBO(m_j)$,
% \begin{equation}
% \begin{split}
%     \arg\min_{m} \Big[||Ex-y||^{2}_{2}-\sum_{x\in\Omega(m)}ELBO(x)\Big]; ELBO(x) = \mathbb{E}_{D_{\theta_d}(z|x)} \Big[\log G_{\theta_g}(x|\textbf{z})+\log\frac{p(z)}{G_{\theta_g}(x|\textbf{z})}\Big].
% \end{split}
% \end{equation}
% These methods have mainly been evaluated on the Human Connectome Project
% (HCP) \citep{van2012human} dataset and have shown good performance on 4x undersampled images (see Fig \ref{fig_DDP}).
%%%%%%%%%%%%%
%%%%%%%%
%%%
%
%%%%%%%%
%%%%
%
% \subsection{Bayesian Learning}
% \label{sec_gen_bayesian}
% Bayes’s theorem expresses the posterior $p(x|y)$ as a function of the k-space data likelihood, $p(y|x)$ and the prior $p(x)$ with the form $p(x|y)\propto p(y|x)p(x)$ also known as the ``product-of-the-experts'' in DL literature. 
% In \citep{tezcan2018mr}, the prior is estimated with a Monte Carlo sampling technique which is computationally intensive. 
% To overcome the computational cost, the method proposed in \citep{luo2020mri} considers a generative regression model called PixelCNN+ \citep{oord2016conditional} to estimate the prior $p(x)$. 
% This method demonstrated very good performance, i.e. they achieved more than 3 dB PSNR improvement than the current state-of-the-art methods like GRAPPA, variational networks (see Sec. \ref{sec_nongen_cnn}) and SPIRIT algorithms. 
% Despite showing good results, the overall computational cost and running time is very high compared to GAN or VAE based methods.
%%%%%%%%%
%%%%
%
\subsection{Active Acquisition Methods}
\label{sec_active_learning}
\begin{figure}
    \centering
    \includegraphics[width=\textwidth]{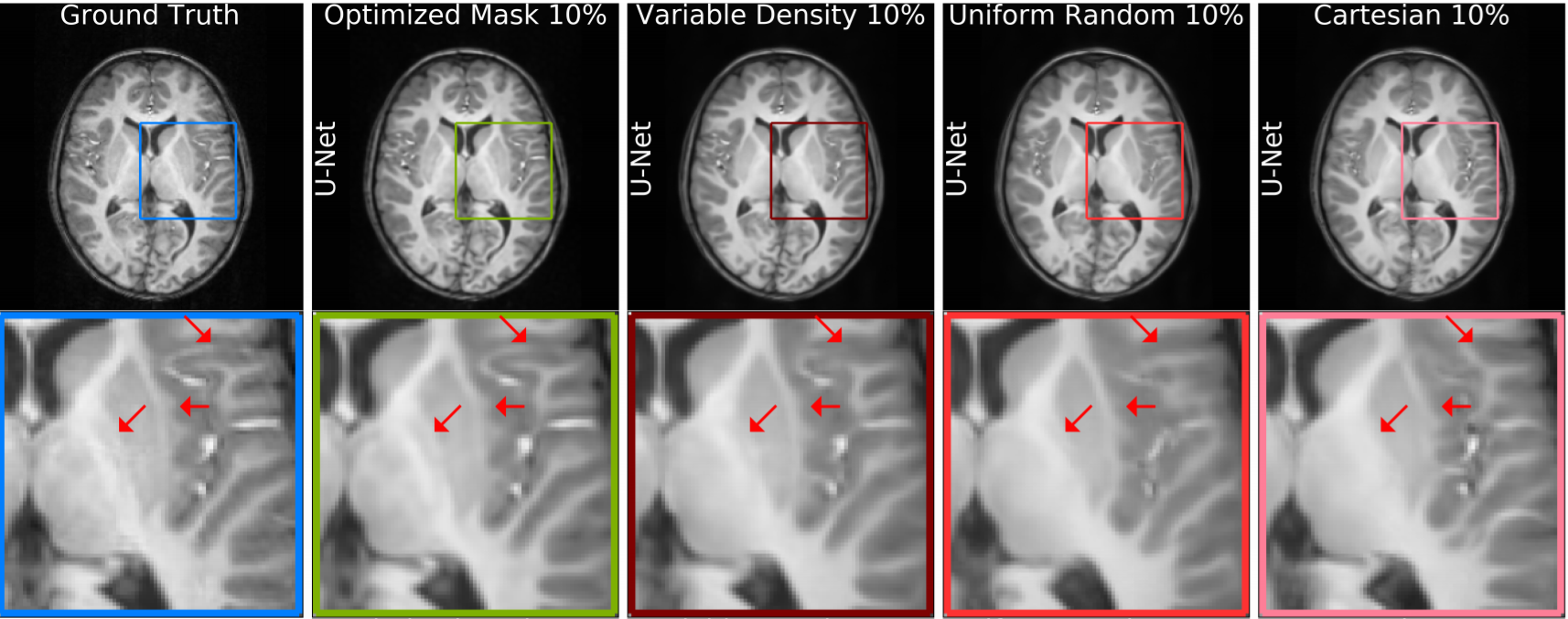}
    \caption{\textbf{MR Reconstruction using Active Acquisition Method}: Variable mask learned by the active acquisition method LOUPE \citep{LOUPE} can improve the image quality of a UNet based MR image reconstruction task. The leftmost image is the ground truth, the second image shows the reconstruction using an optimized mask, the next image is the variaable mask, that is followed by the uniformly random and Cartesian mask. The images are borrowed from the LOUPE \citep{LOUPE} paper.}
    \label{fig:active}
\end{figure}

\noindent{\bf Combined k-space and image methods}: 
All of the above methods consider a fixed  k-space sampling that is predetermined by the user. This sampling process is isolated from the reconstruction pipeline. Recent works have investigated if the sampling process itself can be included as a part of the reconstruction optimization framework. A basic overview of these works can be described as follows:
\begin{itemize}
    \item The algorithm has access to the fully sampled training MR images $\{\textbf{x}_1, \textbf{x}_2, \cdots, \textbf{x}_N\}$
    \item The encoder, $G_{\theta_g}(\cdot)$, learns the sampling pattern by optimizing parameter $\theta_{g}$.
    \item The decoder, $D_{\theta_d}(\cdot)$, is the reconstruction algorithm that is parameterized by $\theta_{d}$
    \item The encoder $G_{\theta_g}(\cdot)$ is optimized by minimizing the empirical risk on the training MR images, $\frac{1}{N}\sum_{i=1}^{N}L_{q}(\textbf{x}_{i}, D_{\theta_d}(G_{\theta_g}(\mathscr{F}(\textbf{x}))))$, where $L_{q}$ is some arbitrary loss of the decoder.
\end{itemize}
This strategy was used in LOUPE \citep{LOUPE} where a network was learnt to optimize the  under-sampling  pattern such that $G_{\theta_g}(\cdot)$ provided a probabilistic sampling mask $\mathcal{M}(\cdot)$ assuming each line in k-space as an independent Bernoulli random variable by optimizing:
\begin{equation}
    \arg\min_{\theta_g, \theta_d}\mathbb{E}_{\mathcal{M}\sim G_{\theta_g}}[||D_{\theta_d}(\mathcal{M}\mathscr{F}\textbf{x}) - \textbf{x}||_{1} + \lambda\mathcal{M}],
\end{equation}
where $D_{\theta_d}$ is an anti-aliasing deep neural network. Experiments on $T_1$-weighted structural brain MRI scans show that the LOUPE method improves PSNR by $\sim 5\%$ with respect to the state-of-the-art methods that is shown in Fig. \ref{fig:active}, second column. 
%%%%%%%%%%%%%
%%%%%%%%%%
%%%%%%
%
%%%%%%%%%%%
%%%%%%%%%
%%%%
%
A follow-up work to LOUPE \citep{DLOUPE} imposed a hard sparsity constraint on $G_{\theta_g}(\cdot)$ to ensure robustness to noise. 
In the deep active acquisition method \citep{Active_acq}, $G_{\theta_g}(\cdot)$ is termed the evaluator and $D_{\theta_d}(\cdot)$ is the reconstruction network. Given a zero-filled MR image,  $D_{\theta_d}(\textbf{x}^{ZF})$ provides the reconstructed image and the uncertainty map. The evaluator $G_{\theta_g}(\cdot)$ decomposes the reconstructed image and ground truth image into spectral maps and provides a score to each k-space line of the reconstructed image. Based on the scores, the methodology decides to acquire the appropriate k-space locations from the MR scanner. 
The Deep Probabilistic Subsampling (DPS) method in \citep{DPS} develops a task-adaptive probabilistic undersampling scheme using a softmax based approach followed by MR reconstruction. 
On the other hand, the work on joint model based deep learning (J-MoDL) \citep{JMODL} 
optimized both sampling and reconstruction  using  Eqns \ref{eq_self_reg} and \ref{eq_self_dc} to jointly optimize a data consistency network and a regularization network. The data consistency network is a residual network that acts as a denoiser, while the regularization network decides the sampling scheme. 
The PILOT \citep{weiss2021pilot} method also jointly optimizes the k-space sampling and the reconstruction. The network has a sub-sampling layer to decide the importance of a k-space line, while the regridding and the task layer jointly reconstruct the image. 
The optimal k-space lines are chosen either using the greedy traveling salesman problem or imposing acquisition machine constraints. 
Joint optimization of k-space sampling and reconstruction also appeared in recent methods such as \citep{ERAKI, JointEstimation}.
%%%%%%%%%%%%
%%%%%
%
\section{Inverse Mapping using Non-generative Models}
\label{sec_nongen}
In this section we discuss non-generative models that use the following optimization framework:
\begin{equation}
\begin{split}
    & \hat{\textbf{x}} = \arg\min_{\textbf{x}}\frac{1}{2}||\textbf{y} - A\textbf{x}||^{2}_{2} + \lambda\mathcal{L}_{\theta_g}; \quad \mathcal{L}_{\theta_g} = \arg\min_{\theta_{g}}\mathbb{E}_{\textbf{x}\sim p_{data}}\frac{1}{2}||\textbf{x}_{j} - G_{NGE}(\textbf{x}|c,\theta_{g})||^{2}_{2}.
\end{split}
\end{equation}
The non-generative models also have a data consistency term and a regularization term similar to Eqn. \ref{eq_3_tikonov}. 
As discussed earlier in section \ref{sec_intro_deep_learning}, however, the non-generative models do not assume any underlying distribution of the data and learn the inverse mapping by parameter optimization using Eqn. \ref{eq_5_non_generative_models}. Below, we discuss the different types of non-generative models.
%%%%%%%%%%%%%%%
%%%%%%%%
%%%
%
\subsection{Perceptron Based Models}
\label{sec_nongen_MLP}
\begin{figure}[!t]
    \centering
    \includegraphics[width=\textwidth]{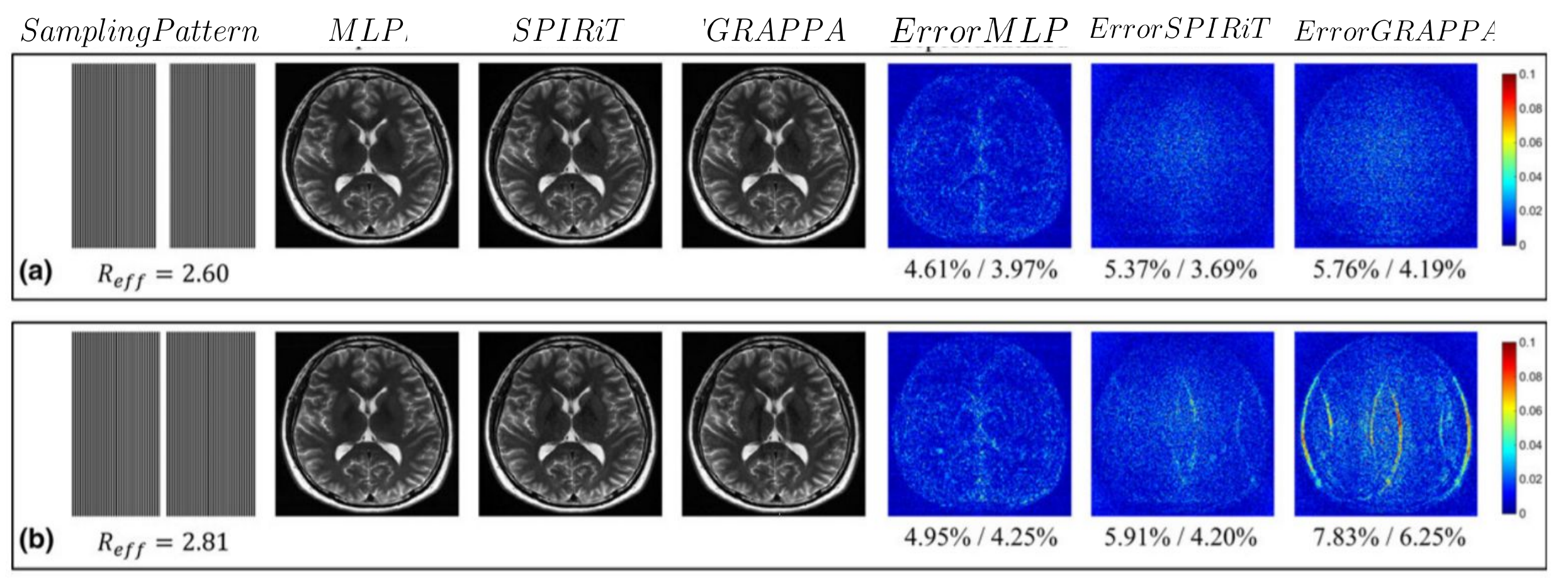}
    \caption{\textbf{Visual Comparison with Perceptron, SPIRiT and GRAPPA methods.} We note the dealiasing capability of the perceptron based method at different sampling rates, i.e. row (a) and row (b), for a $T_2$-weighted image are visibly better than the classical methods. The figure is borrowed from \citep{kwon2017parallel} paper. }
    \label{fig_perceptron}
\end{figure}

%\noindent{\bf Inverse mapping using image space rectification}:
The work in \citep{kwon2017parallel, cohen2018mr} developed a multi-level perceptron (MLP) based learning technique that learns a nonlinear relationship between the k-space measurements, the aliased images, and the desired unaliased images. 
The input to an MLP is the real and  imaginary part of an aliased image and the k-space measurement, and the output is the corresponding unaliased image. 
We show a visual comparison  of this method \citep{kwon2017parallel} with the SPIRiT and GRAPPA methods in Fig. \ref{fig_perceptron}. 
This method showed better performance with lower RMSE at different undersampling factors. 
%%%%%%%%%%%%%%%%%%%
%%%%%%%%%%%%
%%%%%
%%
%
%%%%%%%%%%
%%%%%%
%%%
%
%%%%%%%%%%%%%%%%%%%%%%%
%%%%%%%%%%%%%%
%%%%%%%%
%%%
%
\subsection{Untrained Networks}
\label{sec_nongen_untrained}
So far, we have talked about various deep leraning architectures and their training strategies using a given training dataset. 
The most exciting question one can ask ``is it always necessary to train a DL network to obtain the best result at test time?'', 
or ``can we solve the inverse problem using DL similar to classical methods that do not necessarily require a training phase to learn the parameter priors''? 
We note several state-of-the-art methods that uses ACS lines or other k-space lines of the k-space measurement $y$ to train a DL network instead of an MR image as a ground truth. 
The robust artificial neural network for k-space interpolation (RAKI) \citep{RAKI} trains a CNN by using the ACS lines. 
The RAKI methodology shares some commonality with GRAPPA. 
However, the main distinction is the linear estimation of the convolution kernel in GRAPPA which is replaced with a non-linear kernel in CNN. 
The CNN kernels are optimized using the following objective function:
\begin{equation}
\begin{split}
    & \hat{\textbf{x}} = \arg\min_{\textbf{x}}\frac{1}{2}||\textbf{y} - A\textbf{x}||^{2}_{2} + \lambda\mathcal{L}_{\theta_g}; \quad \loss_{\theta_g} = ||\textbf{y}^{ACS} - G_{NGE}(\Tilde{\textbf{y}}^{ACS}; \theta_g)||^{2}_{F}
\end{split}
\end{equation}
where, $y^{ACS}$ are the acquired ACS lines, and $G_{NGE}(\cdot)$ is the CNN network that performs MRI reconstruction. 
The RAKI method has shown $ 0\%, 0\%, 11\%, 28\%, \text{and}\; 41\%$ improvement in RMSE score with respect to GRAPPA on phantom images at $\{2x, 3x, 4x, 5x, 6x\}$ acceleration factors respectively. 
A followup work called residual RAKI (rRAKI) \citep{zhang2019scan} improves the RMSE score with the help of a residual network structure. 
The LORAKI \citep{kim2019loraki} method is based on the low rank assumption of LORAKS \citep{haldar2013low}. 
It uses a recurrent CNN network to combine the auto-calibrated LORAKS \citep{haldar2013low} and the RAKI \citep{RAKI} methods. 
On five different slices of a $T_2$-weighted dataset, the LORAKI method has shown good improvement in SSIM scores compared to GRAPPA, RAKI, AC-LORAKS among others. 
Later, the sRAKI-RNN \citep{hosseini2019sraki} method proposed a unified framework that performs regularization through calibration and data consistency using a more simplified RNN network than LORAKI.

Deep Image Prior (DIP) and its variants \citep{ulyanov2018deep, BayesianDeep, gandelsman2019double} have shown outstanding results on computer vision tasks such as denoising, in-painting, super resolution, domain translation etc. 
A vanilla DIP network uses a randomly weighted autoencoder, $D_{\theta_d}(G_{\theta_g}(\textbf{z}))$, that reconstructs a clean image $x\in R^{W\times H \times 3}$ given a fixed noise vector $\textbf{z}\in R^{W\times H \times D}$. The network is optimized using the ``ground truth" \textit{noisy} image $\hat{\textbf{x}}$. A {manual or user chosen} ``early stopping'' of the optimization is required as optimization until convergence overfits to noise in the image. 
A recent work called Deep Decoder \citep{heckel2018deep} shows that an under-parameterized decoder network, $D_{\theta_d}(\cdot)$, is not expressive enough to learn the high frequency components such as noise and can nicely approximate the denoised version of the image. 
The Deep Decoder uses pixel-wise linear combinations of channels and shared weights in spatial dimensions that collectively help it to learn relationships and characteristics of
nearby pixels. 
It has been recently understood that such advancements can directly be applied to MR image reconstruction \citep{canuntrained}. 
Given a set of k-space measurements $\textbf{y}_{1}, \textbf{y}_{2}, \cdots, \textbf{y}_{n}$ from receiver coils, an un-trained network $G_{\theta}(\textbf{z})$ uses an iterative first order method to estimate parameters $\hat{\theta}$ by optimizing;
\begin{equation}
    \min\loss_{\theta} = \frac{1}{2}||\textbf{y}_i - \mathcal{M}\mathscr{F}G(\textbf{z};\theta)||^{2}_{2}.
    \label{eq_DIP}
\end{equation}
The network is initialized with random weight $\theta_{0}$ and then optimized using Eqn. \ref{eq_DIP} to obtain $\hat{\theta}$. 
%Corresponding MR image is estimated using the $\hat{\textbf{x}}_{i}=G_{i}(\textbf{z};\hat{\theta})$ and the image $x = \sqrt{\sum_{i=1}^{n}}|\hat{\textbf{x}}_{i}|^{2}$. 
The work in \citep{using_untrained} introduces a feature map regularization: $\frac{1}{2}\sum_{j=1}^{L}||D_{j}\textbf{y}_{i} - \mathcal{M}\mathscr{F}G_{j,i}(\textbf{z};\theta)||^{2}_{2}$, in Eqn. \ref{eq_DIP} where $D_{j}$ matches the features of $j^{th}$ layer. This term encourages fidelity between the network's intermediate representations and the acquired k-space measurements. 
The works in \citep{DIP_theoritical_gurantee, heckel2020compressive} provide theoretical guarantees on recovery of image $x$ from the k-space measurements. 
Recently proposed method termed ``Scan-Specific  Artifact  Reduction  in  k-space" or SPARK \citep{arefeen2021scan} trains a CNN  to estimate and correct k-space errors made by an input reconstruction technique. The results of this method are also quite impressive given that only ACS lines are used for training the CNN. 
Along similar lines, the authors in \citep{yoo2021time} used the Deep Image Prior setup for dynamic MRI reconstruction. 

%%%%%%%%%%%%
%%%%%
%

In the self supervised approach, a subset of the undersampled k-space lines are typically used to validate the DL network in addition to the acquired undersampled k-space lines that are used to optimize the network. 
Work in this direction divides the total k-space lines into two portions:(i) k-space lines for data consistency $\textbf{y}^{dc}$ and (ii) k-space lines $\textbf{y}^{loss}$ for regularization. %The k-space line of $y^{dc}$ are used for data consistency, $||x - G_{\theta_g}(y^{dc})||^{2}_{2}$ and the k-space lines of $y^{loss}$ are used to optimize the loss function. 
%The reconstruction network $G_{\theta_g}(\cdot)$ reconstructs the MR image $x$ given $y^{dc}$ in a self-supervised physics based DL \citep{SelfSupervisedPhysics} framework and the network is optimized using the following loss function:
%\begin{equation}
%    \min_{\theta_g} \loss(y^{loss}, \mathcal{M}^{loss}\mathscr{F}G_{\theta_g}(y^{dc})),
%    \label{eq_self_sup}
%\end{equation}
%where $\mathcal{M}^{loss}$ is the binary mask that collects the $y^{loss}$ k-space lines. 
In \citep{SelfSupMulti}, the authors use a multi-fold validation set of k-space data to optimize the DL network. 
Other methods such as SRAKI \citep{hosseini2020accelerated, hosseini2019accelerated} use the self-supervision to reconstruct the images. 
A deep Reinforcement learning based approach is studied in \citep{SelfSupDeepAcive} that deploys a reconstruction network and an active acquisition network. 
Another method such as \citep{yaman2021zero} provides an unrolled optimization algorithm to estimate the missing k-space lines. Other methods that fall under this umbrella include the transformer based method \citep{korkmaz2021unsupervised} and a scan specific optimization method \citep{yaman2021scan, tamir2019unsupervised}. 
%%%%%%%%%%%%%%%
%%%%%%%%%%%
%%%%%
%%
%

\subsection{Convolutional Neural Networks}
\label{sec_nongen_cnn}
Spatial models are mostly dominated by the various flavours of CNNs such as complex-valued CNN \citep{wang2020deepcomplexmri, cole2019complex}, unrolled optimization using CNN \citep{schlemper2017deep}, variational networks \citep{hammernik2018learning}, etc. 
Depending on how the MR images are reconstructed, we divide all CNN based spatial methods into the following categories.
%%%%%%%%%%%%%%%
%%%%%%%%%%%
%%%%%
%%
%

\begin{figure}[!t]
    \centering
    \includegraphics[width=\textwidth]{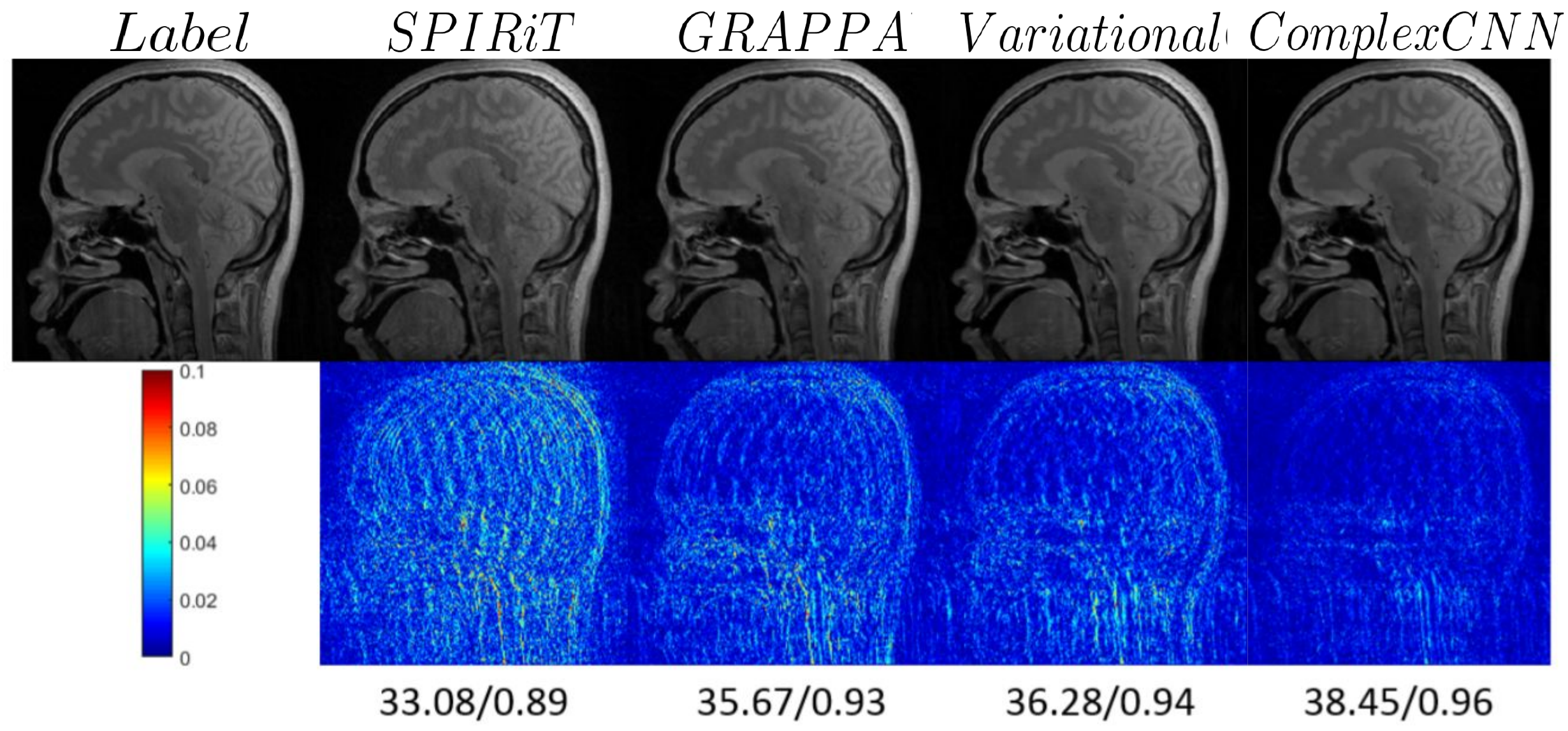}
    \caption{\textbf{Comparison of a Complex-valued CNN with other state-of-the-art methods:} We show visual comparison of SPIRiT \citep{lustig2010spirit}, GRAPPA \citep{bouman1993generalized}, VariationalNet, and the ComplexCNN. We note from the difference map that the ComplexCNN performed well with respect to the other state-of-the-art methods. The PSNR and SSIM values are given at the bottom for each method.}
    \label{fig_deepcomplex}
\end{figure}

\noindent{\bf Inverse mapping from k-space}: 
The AUTOMAP \citep{Automap} learns a reconstruction mapping using a network having three fully connected layers (3 FCs) and two convolutional layers (2 Convs) with an input dimension of $128\times 128$. 
Any image of size more than $128 \times 128$ is cropped and subsampled to $128 \times 128$. 
The final model yielded a PSNR of 28.2 on FastMRI knee dataset outperforming the previous validation baseline of 25.9 on the same dataset. 
Different from these methods, there are a few works \citep{wang2020deepcomplexmri, cole2019complex} that have used CNN networks with complex valued kernels to reconstruct MR images from complex valued k-space measurements. 
The method in \citep{DeepcomplexMRI} uses a complex valued ResNet (that is a type of CNN) network and is shown to obtain good results on 12-channel fully sampled k-space dataset (see Fig. \ref{fig_deepcomplex} for a visual comparison with other methods). 
Another method uses a Laplacian pyramid-based complex neural network \citep{liang2020laplacian} for MR image reconstruction. 
%%%%%%%%%%%%%%%%%%%%%
%%%%%%%%%%%%%%
%%%%%%%
%%
%
%%%%%%%%%%%%%%%
%%%%%%%%%%%
%%%%%
%%
%

\begin{figure}[!t]
    \centering
    \includegraphics[width=\textwidth]{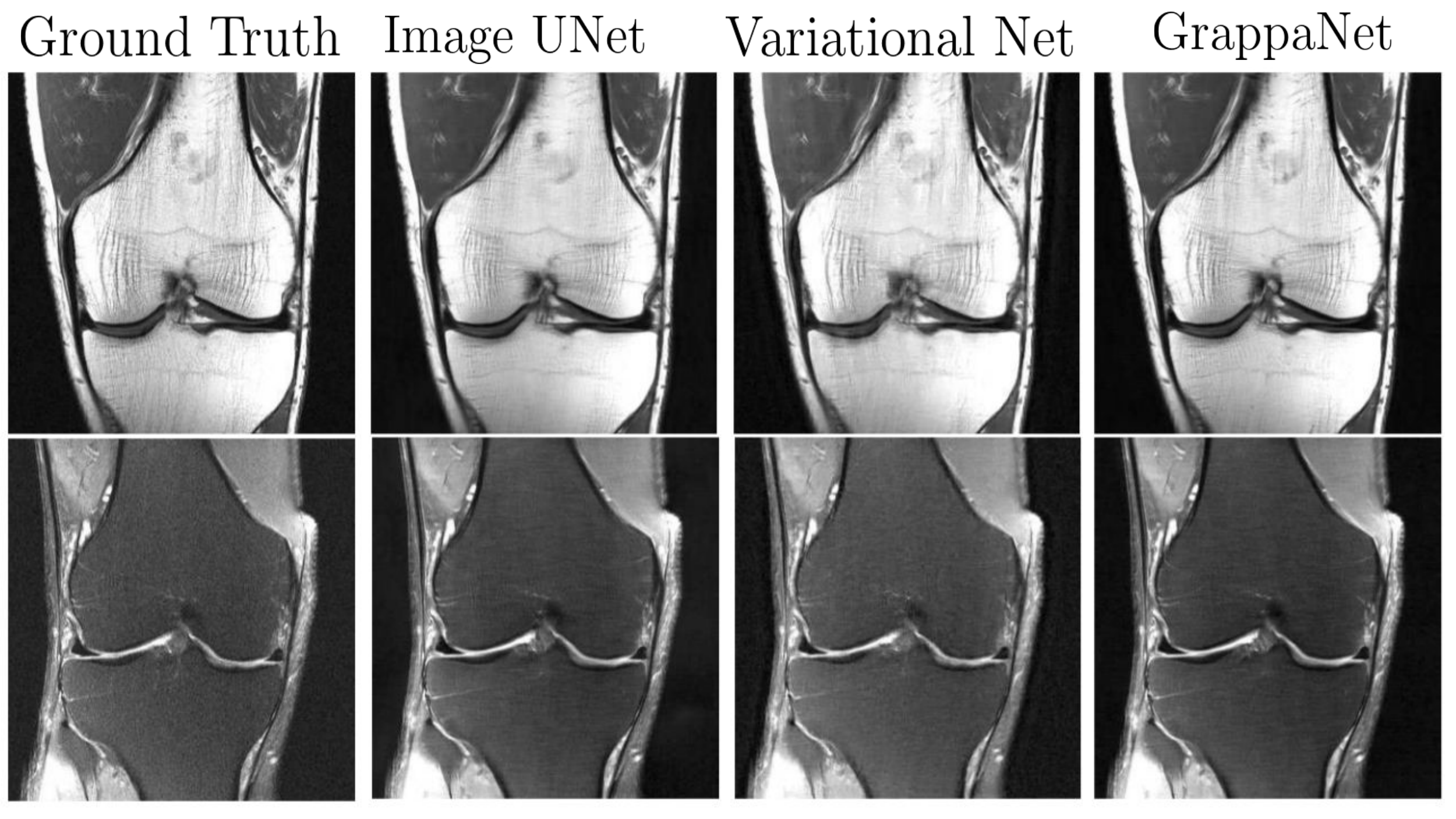}
    \caption{\textbf{GrappaNet Comparison with VariationalNET, UNet methods}}
    \label{fig_grappa}
\end{figure}

\noindent{\bf Inverse Mapping for Image Rectification}: 
%Unrolled optimization based methods in Sec. \ref{sec_classic_image} show that the intermediate variables regularizes in the k-space domain. 
In CNN based sequential spatial models such as DeepADMM net models \citep{sun2016deep, schlemper2017deep} and Deep Cascade CNN (DCCNN) \citep{schlemper2017deep}, the regularization is done in image space using the following set of equations:
\begin{equation}
    \begin{split}
        & \arg\min_{\textbf{x}} \frac{1}{2}||A\textbf{x}-\textbf{y}||^{2}_{2} + ||\textbf{x}+\beta-\textbf{z}||^{2}_{2}; \quad \arg\min_{\textbf{z}_{i}} \sum_{i}[\lambda_{i}g(\Gamma_{i}\textbf{z}_{i})||\textbf{x}+\beta-\textbf{z}||^{2}_{2}]\\
        & \beta \leftarrow \beta + \alpha(\textbf{x} - \textbf{z}).
    \end{split}
\end{equation}
Here, $\alpha$ and $\beta$ are Lagrange multipliers. 
The ISTA net \citep{zhang2018ista} modifies the above image update rule as follows, $\textbf{x}_{i} = \arg\min_{\textbf{x}}\frac{1}{2}||C(\textbf{x}) - C(\textbf{z}_{i})||^{2}_{2} + \lambda||C(\textbf{x})||_{1}$, using a CNN network $C(\cdot)$. 
%We note that, the DeepADMM \citep{sun2016deep} has attained $\sim 41.37$ PSNR on a private Brain dataset. 
Note that the DeepADMM network demonstrated good performance when the network was trained on brain data but tested on chest data. 
Later, MODL \citep{MODL} proposed a model based MRI reconstruction where they used a convolution neural network (CNN) based regularization prior. 
Later, a dynamic MRI using MODL based deep learning was proposed by \citep{biswas2019dynamic}. 
The optimization, i.e. $\arg\min_{\textbf{x}} \frac{1}{2}||A\textbf{x}-\textbf{y}||^{2}_{2} + \lambda ||C(\textbf{x})||^{2}_{2}$, denoises the alias artifacts and noise using a CNN network $C(\cdot)$ as a regularization prior, and $\lambda$ is a trainable parameter.
%One major drawback of these networks is that, the underlying physics of k-space acquisition process is not captured within the optimization framework of CNN network. 
To address this concern, a full end-to-end CNN model called GrappaNet \citep{sriram2020grappanet} was developed, which is a nonlinear version of GRAPPA set within a CNN network. 
The CNN network has two sub networks; the first sub network, $f_{1}(\textbf{y})$, fills the missing k-space lines using a non-linear CNN based interpolation function similar to GRAPPA. 
Subsequently, a second network, $f_{2}$, maps the filled k-space measurement to the image space. 
The GrappaNet model has shown excellent performance ($40.74$ PSNR, $0.957$ SSIM) on the FastMRI dataset and is one of the best performing methods. 
A qualitative comparison is shown in Fig. \ref{fig_grappa}.
%%%%%%%%%%%%%%
%%%%%%%%%%
%%%%
%
%%%%%%%%%%%
%%%%%%%%%
%%%
%
Along similar lines, a deep variational network \citep{hammernik2018learning} is used to MRI reconstruction.  
Other works, such as \citep{Accerleraed, cheng2018highly, aggarwal2018modl} train the parameters of a deep network by minimizing the reconstruction error between the image from zero-filled k-space and the image from fully sampled k-space. 
The cascaded CNN network learns spatio-temporal correlations efficiently by combining convolution and data sharing approaches in \citep{schlemper2017deep}. 
The \citep{seegoolam2019exploiting} method proposed to use a CNN network to 
estimate motions from undersampled MRI sequences that is used to fuse data along the entire temporal axis. 

%%%%%%%%%%%%%%
%%%%%%%%
%%%%%%
%%
%
\subsection{Recurrent Neural Networks}
\label{sec_nongen_rnn}
\noindent{\bf Inverse mapping from k-space}: 
We note that a majority of the iterative temporal networks, a.k.a the recurrent neural network models, are k-space to image space reconstruction methods and typically follow the optimization described in Section \ref{sec_nongen_cnn}. 
The temporal methods, by design, are classified into two categories, namely (i) regularization methods, and (ii) variable splitting methods. 

Several state-of-the-art methods have considered temporal methods as a way to regularize using the iterative hard threshold (IHT) method from  \citep{blumensath2009iterative} that approximates the $l_{0}$ norm. 
Mathematically, the IHT update rule is as follows:
\begin{equation}
\label{eq_iht}
    \textbf{x}_{t+1} = H_{k}[\textbf{x}_{t} - \alpha \Phi^{T}(\Phi \textbf{x}_{t}-\textbf{y})], 
\end{equation}
where $\alpha$ is the step-size parameter, $H_{k}[\cdot]$ is the operator that sets all but $k$-largest values to zero (proxy for $l_{0}$ operation), and the dictionary $\Phi$ satisfies the restricted isometry property (RIP)\footnote{\textbf{Restricted Isometry Property (RIP):} The projection from the measurement matrix $E$ in Eqn. \ref{eq_6_compressed_sensing} should preserve the distance between two MR images $\textbf{x}_1$ and $\textbf{x}_2$ bounded by factors of $1-\delta$ and $1+\delta$, i.e. $(1-\delta)||\textbf{x}_1 - \textbf{x}_2||^{2}_{2}\leq||E(\textbf{x}_1 - \textbf{x}_2)||^{2}_{2}\leq (1+\delta)||\textbf{x}_1 - \textbf{x}_2||^{2}_{2}$, where $\delta$ is a small constant.}. 
The work in \citep{xin2016maximal} shows that this hard threshold operator resembles the memory state of the LSTM network. Similar to the clustering based sparsity pattern of IHT, the gates of LSTM inherently promotes sparsity. 
Along similar lines, the Neural Proximal Gradient Descent work \citep{mardani2018neural} envisioned a one-to-one correspondence between the proximal gradient descent operation and the update of a RNN network. 
Mathematically, an iteration of a proximal operator $P_f$ given by: $\textbf{x}_{t+1} = P_{f}(\textbf{x}_{t} + \alpha\Phi^{H}(\textbf{y}-\Phi \textbf{x}))$, resembles the LSTM update rule:
\begin{equation}
    \begin{split}
        & \textbf{s}_{t+1} = g(\textbf{x}_{t}; \textbf{y}); \quad \textbf{x}_{t+1} = P_{f}(\textbf{s}_{t+1}),
    \end{split}
\end{equation}
where $g(\textbf{x}_{t}; \textbf{y}) = \textbf{x}_{t} + \alpha\Phi^{H}(\textbf{y}-\Phi \textbf{x})$ is the update step, $\textbf{s}_{t+1}$ is the hidden state and $\Phi$ is a dictionary. 
Different from these, a local-global recurrent neural network is proposed in \citep{guo2021over} that uses two recurrent networks, one network to capture high frequency components, and another network to capture the low frequency components. 
The method in \citep{oh2021k} uses a bidirectional RNN and replaces the dense network structure of  \citep{Automap} while removing aliasing artifacts in the reconstructed image. 

The Convolutional Recurrent Neural Networks or CRNN \citep{qin2018convolutional} method proposed a variable splitting and alternate minimisation method using a RNN based model. 
Recovering finer details was the main challenge of the PyramidRNN \citep{wang2019pyramid} that proposed to reconstruct images at multiple scales. 
Three CRNNs are deployed to reconstruct images at different scales, i.e. $\textbf{x}^{0} = CRNN(\textbf{x}^{zero}, \textbf{y}, \theta_{1}), \textbf{x}^{1} = CRNN(\textbf{x}^{0}, \textbf{y}, \theta_{2}), \textbf{x}^{2} = CRNN(\textbf{x}^{1}, \textbf{y}, \theta_{3})$, and the final data consistency is performed after $\textbf{x}^{0}, \textbf{x}^{1}, \textbf{x}^{2}$ are combined using another CNN. 
The CRNN is used as a recurrent neural network in the variational approach of VariationNET \citep{sriram2020end}, i.e. 
\begin{equation}
    \textbf{x}_{t+1} = \textbf{x}_{t} - \alpha\Phi^{H}(\textbf{y}-\Phi \textbf{x}) + D(\textbf{x}_{t}),
\end{equation}
where $D(\cdot)$ is a CRNN network that provides MR reconstruction. 
In this unrolled optimization method, the CRNN is used as a proximal operator to reconstruct the MR image. 
The VariationNET is a followup work of Deep Variational Network of \citep{hammernik2018learning} that we discussed in Sec. \ref{sec_nongen_cnn}. 
The VariationalNET unrolls an iterative algorithm involving a CRNN based recurrent neural network based regularizer, while the Deep Variational Network of \citep{hammernik2018learning} unrolls an iterative algorithm involving a receptive field based convolutional regularizer.

%%%%%%%%%%
%%%%
%
\subsection{Hypernetwork Models}
\label{sec_hypernetworks}
Hypernetworks are meta-networks that regress optimal weights of a task network (often called as data network \citep{ZSTT} or main network \citep{Hypernetworks}). The data network $G_{NGE}(\textbf{x}^{low}, \theta_g)$ performs the mapping from aliased or the low resolution images $\textbf{x}^{low}$ to the high resolution MR images $\textbf{x}^{gen}$. The hypernetwork $H(\alpha, \theta_{hyp})$ estimates weights $\theta_g$ of the network $G_{NGE}(\theta_g)$ given the random variable $\alpha$ sampled from a prior distribution $\alpha\sim p(\alpha)$. The end-to-end network is trained by optimizing:
\begin{equation}
    \arg\min_{\psi} \mathbb{E}_{\theta_g\sim H(\alpha, \theta_{hyp})}[||G_{NGE}(\textbf{x}^{low}, \theta_g) - \textbf{x}||_{2}^{2} + \mathcal{R}(G_{NGE}(\textbf{x}^{low}, \theta_g))].
\end{equation}
In \citep{HyperMRI}, the prior distribution $p(\alpha)$ is a uniform distribution $\mathcal{U}[-1,+1]$ (and the process is called Uniform Hyperparameter Sampling) or the sampling can be based on the data density (called data-driven hyperparameter sampling). Along similar lines, the work in \citep{MAC} trained a dynamic weight predictor (DWP) network that provides layer wise weights to the data network. The DWP generates the layer wise weights given the context vector $\gamma$ that comprises of three factors such as the anatomy under study, undersampling mask pattern and the acceleration factor.
%%%%%%%%%%
%%%%%
%

\section{Comparison of state-of-the-art methods}
\label{sec_experiments}
%%%%%%%%%%%%%%%%%%
%%%%%%%%%
%
\begin{table*}[!t]
\resizebox{\textwidth}{!}{%
\begin{tabular}{|c|l|l|l|l|c|l|l|l|l|}
\hline
\multicolumn{1}{|c|}{Acceleration} & Model & \multicolumn{1}{c|}{NMSE} & \multicolumn{1}{c|}{PSNR} & \multicolumn{1}{c|}{SSIM} & Acceleration & Model & NMSE & PSNR & SSIM \\ \hline\hline
\multirow{8}{*}{4-fold} & Zero Filled & 0.0198 & 32.51 & 0.811 & \multirow{8}{*}{8-fold} & Zero Filled & 0.0352 & 29.60 & 0.642 \\   
 & SENSE \citep{pruessmann1999sense} & 0.0154 & 32.79 & 0.816 &  & SENSE \citep{pruessmann1999sense} & 0.0261 & 31.65 & 0.762 \\ 
 & GRAPPA \citep{griswold2002generalized} & 0.0104 & 27.79 & 0.816 &  & GRAPPA\citep{griswold2002generalized} & 0.0202 & 25.31 & 0.782 \\ 
 & RfineGAN\citep{RefineGAN} & 0.0138 & 34.00 & 0.901 &  & RefineGAN\citep{RefineGAN} & 0.0221 & 32.09 & 0.792 \\  
 & DeepADMM \citep{sun2016deep} & 0.0055 & 34.52 & 0.895 &  & DeepADMM\citep{sun2016deep} & 0.0201 & 36.37 & 0.810 \\ 
 & LORAKI \citep{kim2019loraki} & 0.0091 & 35.41 & 0.871 &  & LORAKI \citep{kim2019loraki} & 0.0181 & 36.45 & 0.882 \\ 
 & VariationNET \citep{sriram2020end} & 0.0049 & 38.82 & 0.919 &  & VariationNET \citep{sriram2020end} & 0.0211 & 36.63 & 0.788 \\ 
 & GrappaNet \citep{sriram2020grappanet} & 0.0026 & 40.74 & 0.957 &  & GrappaNet \citep{sriram2020grappanet} & 0.0071 & 36.76 & 0.922 \\  
 & J-MoDL \citep{JMODL} & 0.0021 & 41.53 & 0.961 &  & J-MoDL \citep{JMODL} & 0.0065 & 35.08 & 0.928 \\
 & Deep Decoder \citep{heckel2018deep} & 0.0132 & 31.67 & 0.938 &  & Deep Decoder \citep{heckel2018deep} & 0.0079 & 29.654 & 0.929 \\
 & DIP \citep{ulyanov2018deep}  & 0.0113 & 30.46 & 0.923 &  & DIP \citep{ulyanov2018deep} & 0.0076 & 29.18 & 0.912 \\
  \hline\hline
\end{tabular}%
}
\caption{Normalized mean squared error (NMSE), PSNR and SSIM over the test data with 1000 samples using eight different methods at 4-fold and 8-fold acceleration factors.}
\label{table:qualitative_result}
\end{table*}
%%%%%%%%%%%%%%%%%%
%%%%%%%%%
%
Given the large number of DL methods being proposed, it is imperative to compare these methods on a standard publicly available dataset. 
Many of these methods have shown their effectiveness on various real world datasets using different quantitative metrics such as SSIM, PSNR, RMSE, etc.
There is, however, a scarcity of qualitative and quantitative comparison of these methods on a single dataset. While the fastMRI challenge allowed comparison of several methods, yet, several recent methods from the categories discussed above were not part of the challenge.
Consequently, we compare a few representative MR reconstruction methods both qualitatively and quantitatively on the fastMRI knee dataset \citep{zbontar2018fastmri}. We note that, doing a comprehensive comparison of all the methods mentioned in this review is not feasible due to non-availability of the code as well as the sheer magnitude of the number of methods (running into hundreds).
 We compared the following representative models:
\begin{itemize}
    \item Zero filled image reconstruction method
    \item Classical image space based SENSE method \citep{pruessmann1999sense}
    \item Classical k-space based GRAPPA method \citep{griswold2002generalized}
    \item Unrolled optimization based method called DeepADMM \citep{sun2016deep}
    \item Low rank based LORAKI \citep{kim2019loraki}
    \item Generative adversarial network based RefineGAN \citep{RefineGAN} network
    \item Variational network called VariationNET \citep{sriram2020end}
    \item The deep k-space method GrappaNet \citep{sriram2020grappanet}
    \item Active acquisition based method J-MoDL \citep{JMODL}
    \item Untrained network model Deep Decoder \citep{heckel2018deep} and 
    \item Deep Image Prior DIP \citep{ulyanov2018deep} method.
\end{itemize}
The fastMRI knee dataset consists of raw k-space data from 1594 scans acquired on four different MRI machines. 
We used the official training, validation and test data split in our experiments. 
We did not use images with a width greater than 372 and we note that such data is only $7\%$ of the training data split.
Both the 4x and 8x acceleration factors were evaluated. 
%%%%%%%%%%%%%%%%
%%%%%%%%%%%
%%%%%%%
%%
%
\begin{figure*}[!t]
    \centering
    \includegraphics[ width=\textwidth]{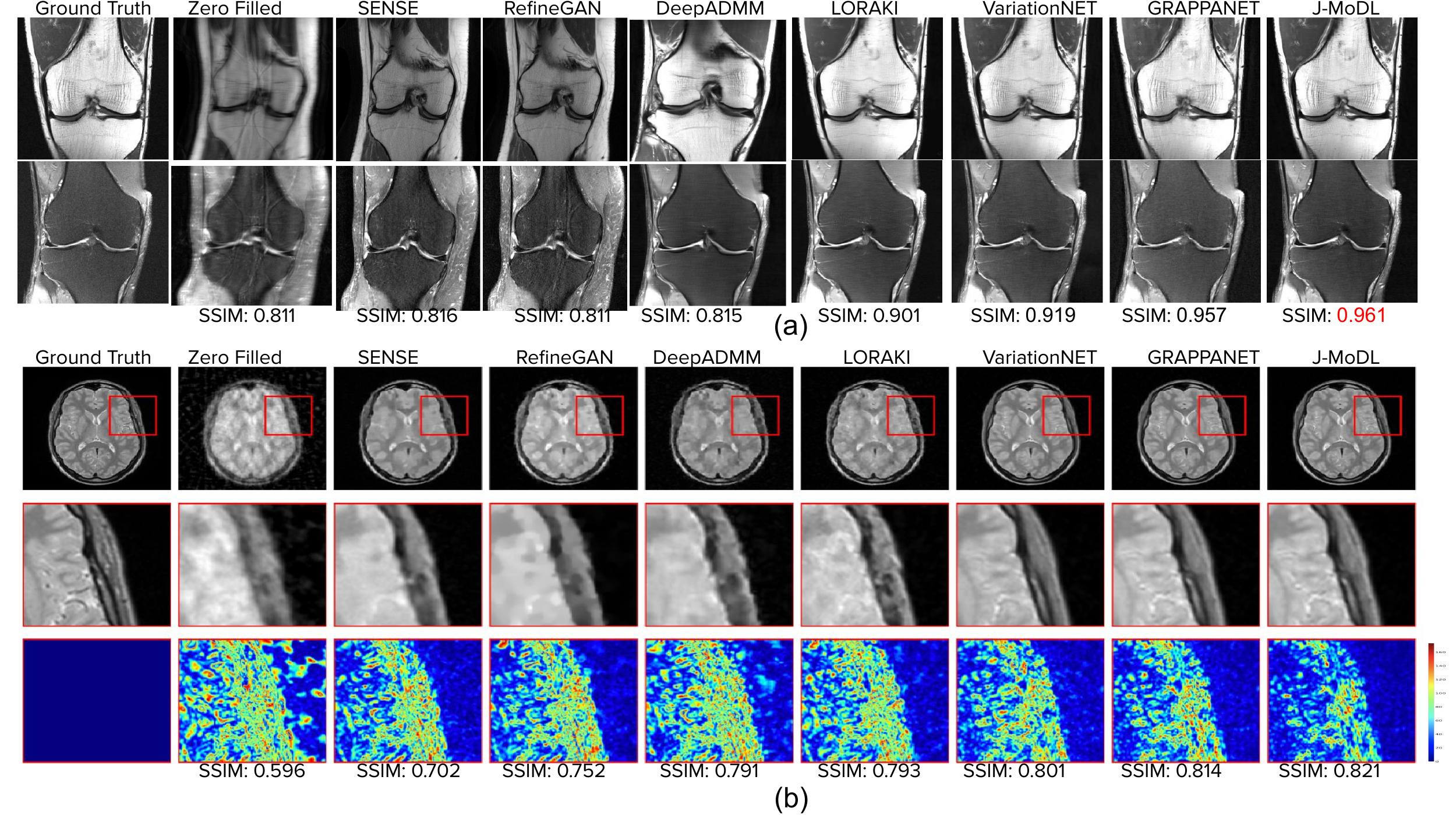}
    \caption{\textbf{Qualitative Comparison:} Comparison for 8x acceleration using ``Zero Filled", SENSE, DeepADMM \citep{sun2016deep}, LORAKI \citep{kim2019loraki}, RefineGAN \citep{RefineGAN}, VariationNET \citep{sriram2020end}, GrappaNet \citep{sriram2020grappanet}, J-MoDL \citep{JMODL} results on fastMRI dataset. We show qualitative results for \textbf{(a)} knee and \textbf{(b)} brain datasets and also report the corresponding SSIM scores.}
    \label{fig:my_label}
\end{figure*}
%%%%%%%%%%%%%%%%
%%%%%%%%%%%
%%%%%%%
%%
%

We used the original implementation
\footnote{
Below are the official implementations of various methods we discussed:\\
\textbf{VariationaNET}: https://github.com/VLOGroup/mri-variationalnetwork/\\
\textbf{GrappaNET}: https://github.com/facebookresearch/fastMRI.git\\
\textbf{RefineGAN}: https://github.com/tmquan/RefineGAN.git\\
\textbf{DeepADMM}: https://github.com/yangyan92/Deep-ADMM-Net.git\\
\textbf{SENSE, GRAPPA}: https://mrirecon.github.io/bart/\\
\textbf{Deep Decoder}: https://github.com/MLI-lab/ConvDecoder.git
} 
of GrappaNet, VariationaNET, SENSE, GRAPPA, DeepADMM, Deep Decoder, and DIP method. 
Similar to GrappaNET, we always use the central 30 k-space lines to compute the training target. 
Treating the real and imaginary as two distinct channels, we dealt with the complex valued input, i.e. we have 30 channel input k-space measurements for the 15-coil complex-valued k-space. 
Where applicable, the models were trained with a linear combination of $L_{1}$ and SSIM loss, i.e.
\begin{equation}
    J(\hat{\textbf{x}},\textbf{x})= -SSIM(G(\textbf{y}),\textbf{x}) + \lambda ||G(y) - x||_{1}
\end{equation}
where $\lambda$ is a hyperparameter, $G(\textbf{y})$ is the model prediction, and $\textbf{x}$ is the ground truth. 

Quantitative results are shown in Table \ref{table:qualitative_result} for several metrics such as NMSE, PSNR, and SSIM scores. 
We observe that GrappaNET, J-MoDL and VariationNET outperformed the baseline methods by a large margin. 
We note that the zero-filled and SENSE reconstructions in Fig \ref{fig:my_label} (a), (b) show a large amount of over-smoothing. 
The reconstruction of SENSE and zero-filled model also lack a majority of the high frequency detail that is clinically relevant, but fine details are visible in case of GrappaNET, VariationNET, J-MoDL, and RefineGAN methods. 
The comparable performance of Deep Decoder and DIP advocates the importance of letting untrained neural network figure out how to perform k-space to MR image reconstruction. 
The J-MoDL method makes heavy use of training data and the joint optimization of k-space lines and the reconstruction of MR images to get good results both for $4\times$ and $8\times$ as shown in Table \ref{table:qualitative_result}. 
On the other hand, the Deep Decoder and DIP methods achieve good performance using untrained networks as discussed in Sec. \ref{sec_nongen_untrained}, which is advantageous as it generalizes to any MR reconstruction scenario.

\section{Discussion}
\label{sec_discussion}
In this paper, we discussed and reviewed several classical reconstruction methods, as well as deep generative and non-generative methods to learn the inverse mapping from k-space to image space. 
Naturally, one might ask the following questions given the review of several papers above: ``are DL methods free from errors?'', ``do they always generalize well?'', and ``are they robust?''.  
To better understand the above mentioned rhetorical questions, we need to discuss several aspects of the performance of these methods such as (i) correct reconstruction of minute details of pathology and anatomical structures; (ii) risk quantification; (iii) robustness; (iv) running time complexity; and (v) generalization.  

Due to the blackbox-like nature of DL methods, the reliability and risk quantification associated with them are often questioned. 
In a recent paper on ``risk quantification in Deep MRI reconstruction'' \citep{edupuganti2020risk}, the authors strongly suggest for quantifying the risk and reliability of DL methods and note that it is very important for accurate patient diagnoses and real world deployment. The paper also shows how Stein’s Unbiased Risk Estimator (SURE) \citep{SURE} can be used as a way to assess uncertainty of the DL model:
\begin{equation}
    SURE = -n\sigma^{2} + ||\hat{\textbf{x}} - \textbf{x}||^{2} + \sigma^{2} trace(\frac{\partial \hat{\textbf{x}}}{\partial \textbf{x}}),
\end{equation}
where the second term represents the end-to-end network sensitivity to small input perturbations. 
This formulation works even when there is no access to the ground truth data $\textbf{x}$. 
In this way, we can successfully measure the risk associated with a DL model. 
In addition to the SURE based method, there are a few other ways to quantify the risk and reliability associated with a DL model. 
The work ``On instabilities of deep learning in image reconstruction'' \citep{antun2019instabilities} uses a set of pretrained models such as AUTOMAP \citep{Automap}, DAGAN \citep{DAGAN}, or Variational Network \citep{sriram2020end} with noisy measurements to quantify their stability.  
This paper as well as several others \citep{narnhofer2021bayesian, antun2019instabilities} have discussed how the stability of the reconstruction process is related to the network architecture, training set and also the subsampling pattern. 

Whether a DL method can capture high frequency components or not is also another area of active research in MR reconstruction.
The robustness of a DL based MR reconstruction method is also studied in various papers such as \citep{pmlrv119raj20a, Addressing_false_negative, caliva2020adversarial, Instabilities}. 
For example, the \citep{Addressing_false_negative, caliva2020adversarial, pmlrv121cheng20a} works perceived adversarial attack as a way to capture minute details during sMR reconstruction thus showing a significant improvement in robustness compared to other methods.  
The work proposed to train a deep learning network with a loss that pays special attention to small anatomical details. 
The methodology progressively adds minuscule perturbation $\delta$ to the input $x$ not perceivable to human eye but may shift the decision of a DL system. 
The method in \citep{pmlrv119raj20a} uses a generative adversarial framework that entails a perturbation generator network $G(\cdot)$ to add minuscule distortion on k-space measurement $y$. 
The work in \citep{Instabilities} proposed to incorporate fast gradient based attack on a zero filled input image and train the deep learning network not to deviate much under such attack. 
The FINE \citep{zhang2020fidelity} methodology, on the other hand, has used pre-trained recon network that was fine-tuned using data consistency to reduce generalization error inside unseen pathologies. 
Please refer to the paper \citep{darestani2021measuring} which provides a summary of robustness of different approaches for image reconstruction.

%In \citep{Addressing_false_negative, caliva2020adversarial} literature, the adversarial attack is perceived as the false negative details those are present in the ground truth but not present in the reconstruction. If a DL model unable to reconstruct a small The paper hypothesized that either those small details lost during undersampling process of MRI, or those small details are not lost but attenuated during process. The method proposes two perturbation techniques on input image \undersampled i.e. \textit{random addition} of small details on \undersampled image and note whether the deep learning model can reconstruct that small details; \textit{finite difference} where the gradient is observed to add small details. The work proposed to train a deep learning network with a loss that leverage special attention mask on small details. 

Optimizing a DL network is also an open area of active research. The GAN networks suffer from a lack of proper optimization of the structure of the network \citep{GAN}. On the other hand, the VAE and Bayesian methods suffer from the large training time complexities. 
We see several active research groups and papers \citep{salimans2016improved, bond2021deep} in Computer Vision and Machine Learning areas pondering upon these questions and coming up with good solutions. 
Also, the work in \citep{l2ornotl2} has shown the effectiveness of capturing perceptual similarity using the $l_2$ and SSIM loss to include local structure in the reconstruction process. Recently, work done in \citep{perpendicular} shows that the standard $l_{2}$ loss prefers or biases the reconstruction with lower image magnitude and hence they propose a new loss function between ground truth $\textbf{x}^{gt}$ and the reconstructed complex-valued images $\textbf{x}$, i.e. $L_{perp}= P(\textbf{x},\textbf{x}^{gt}) + l_1(\textbf{x},\textbf{x}^{gt})$ where $P(\textbf{x},\textbf{x}^{gt}) = |\text{Real}(\textbf{x})\text{Imaginary}(\textbf{x}^{gt}) - \text{Real}(\textbf{x}^{gt})\text{Imaginary}(\textbf{x})|/|\textbf{x}^{gt}|$ which favours the finer details during the reconstruction process. The proposed loss function achieves better performance and faster convergence on complex image reconstruction tasks. 

Regarding generalization,  we note that some DL based models have shown remarkable generalization capabilities, for example: 
the AUTOMAP work was trained on natural images but generalized well for MR reconstruction. However, current hardware (memory) limitations preclude from using this method for high resolution MR reconstruction. On the other hand, some of the latest algorithms that show exceptional performance on the knee dataset have not been extensively tested on other low SNR data. In particular, these methods also need to be tested on quantitative MR modalities to better assess their performance. 

Another bottleneck for using these DL methods is the large amount of training data required. While GAN and Bayesian networks produce accurate reconstruction of minute details of anatomical structures if sufficient data are available at the time of training, it is not clear as to how much training data is required and whether the networks can adapt quickly to change in resolution and field-of-view.
Further, these works have not been tested in scenarios where changes in MR acquisition parameters such as relaxation time (TR), echo time (TE), spatial resolution, number of channels, and undersampling pattern are made at test time and are different from the training dataset.

Most importantly, MRI used for diagnostic purposes should be robust and accurate in reconstructing images of pathology (major and minor). While training based methods have demonstrated their ability to reconstruct normal looking images, extensive validation of these methods on pathological datasets is needed for adoption in clinical settings. To this end, the MR community needs to collaborate and collect normative and pathological datasets for testing. We specifically note that, the range of pathology can vary dramatically in the anatomy being imaged (e.g., the size, shape, location and type of tumor). Thus, extensive training and unavailability of pathological images may present significant challenges to methods that are data hungry. In contrast, untrained networks may provide MR reconstruction capabilities that are significantly better than the current state-of-the-art but do not perform as well as highly trained networks (but generalize well to unknown scenarios). 

Finally, given the exponential rate at which new DL methods are being proposed, several standardized datasets with different degrees of complexity, noise level (for low SNR modalities) and ground truth availability are required to perform a fair comparison between methods. Additionally, fully-sampled raw data (with different sampling schemes) needs to be made available to compare for different undersampling factors. Care must be taken not to obtain data that have already been ``accelerated" with the use of standard GRAPPA-like methods, which might bias the results \citep{Subtel}.

Nevertheless, recent developments using new DL methods point to the great strides that have been made in terms of data reconstruction quality, risk quantification, generalizability and reduction of running time complexity. We hope that this review of DL methods for MR image reconstruction will give researchers a unique viewpoint and summarize in succinct terms the current state-of-the-art methods. We however humbly note that, given the large number of methods presented in the literature, it is impossible to cite and categorize each one of them. As such, in this review, we collected and described broad categories of methods based on the type of methodology used for MR reconstruction.

%Similarly, the VAE, active acquisition and RNN based networks are time efficient, robust and generalize well in real world scenarios. 
%The deep unrolled optimization methods also provide good MR reconstruction and can be trained quickly and running time complexity aspects,

%Recently proposed methods such as SAGAN, DRGAN, AUTOMAP, and GrappaNet collectively produce good qualitative and quantitative results. 
%We also note that, the are recent works those collectively working on various other aspects such as risk quantification, generalizability, reduction of running time complexity, etc., and providing us more sophisticated and robust DL models. We hope that this overview of DL methods for MR image reconstruction will give researchers more insight to the filed and progress the research in this direction.  

%%%%%%%%%%
%%%%
%
%%%%%%%%%%%%%%%%%%%%%%%%%%%%%%%%%%%%%%%%%%%%%%%%%%%%%%%%%%%%%%%%%%%%%%%
% Mandatory Sections. Please complete, especially for final publication
%%%%%%%%%%%%%%%%%%%%%%%%%%%%%%%%%%%%%%%%%%%%%%%%%%%%%%%%%%%%%%%%%%%%%%%

% Acknowledgements.
% Please include any funding, intellectual contributions not included in the authorship, and any other acknowledgements.
\acks{This work was supported by NIH grant: R01MH116173 (PIs: Setsompop, Rathi).}
% Ethical Standards.
% Please edit with the appropriate ethics considerations for your work. Include any pertinent IRB information, etc.
%
% Please note that the submission requirements included:
% The work presented must follow appropriate ethical standards in conducting research and writing the manuscript, following all applicable laws and regulations regarding treatment of animals or human subjects.
\ethics{This work used data from human subjects that is openly available (fastMRI) and was acquired following all applicable regulations as required by the local IRB.}

% Conflict of Interest
% Declaration of possible conflicts of interest: Authors must disclose any financial, organisational, commercial or personal conflicts of interest that might bias their work.
% If no conflicts, please say "We declare we don't have conflicts of interest."
\coi{None.}
{\small
\bibliography{melba-sample}
}

%\newpage
%\subfile{sections/rebuttal}

\end{document}